\documentclass[twocolumn,twocolappendix]{aastex631}
\usepackage{amsmath}
\usepackage{xcolor}

\begin{document}

\title{Describing the Nonuniversal Galaxy Merger Timescales in IllustrisTNG: Effects of Host Halo Mass, Baryons, and Sample Selection}

\correspondingauthor{Kun Xu, Y.P. Jing}
\email{kunxu@sas.upenn.edu; ypjing@sjtu.edu.cn}

\author[0000-0002-7697-3306]{Kun Xu}
\affiliation{Center for Particle Cosmology, Department of Physics and Astronomy,
University of Pennsylvania, Philadelphia, PA 19104, USA}

\author[0000-0002-4534-3125]{Y.P. Jing}
\affil{State Key Laboratory of Dark Matter Physics, Tsung-Dao Lee Institute \& School of Physics and Astronomy, Shanghai Jiao Tong University, Shanghai 201210, People’s Republic of China}



\begin{abstract}
Galaxy merger timescales are crucial for understanding and modeling galaxy formation in our hierarchically structured Universe. However, previous studies have reported widely varying dependencies of merger timescales on initial orbital parameters and mass ratio at the first crossing of $r_{\rm vir}$. Using \texttt{IllustrisTNG} simulations, we find that these dependencies vary with host halo mass, suggesting that discrepancies in prior studies may arise from differences in the systems analyzed. Specifically, in low-mass halos, merger timescales show a stronger dependence on initial orbital parameters, while in high-mass halos, this dependence weakens. To account for these variations, we present a fitting formula that incorporates host mass dependence, achieving a logarithmic scatter smaller than 0.15 dex. Comparing dark matter-only and baryonic simulations, we observe similar merger timescales for circular orbits but notable differences for radial orbits. In halos with $M_{\rm{host}} < 10^{12.5} h^{-1} M_{\odot}$, mergers in dark matter-only runs take longer than in baryonic runs, whereas the trend reverses in more massive halos. We attribute these differences to the competing effects of tidal disruption by central galaxy disks and the resistance of baryonic satellites to tidal stripping. Finally, we extend our model to predict merger timescales from any starting radius within the halo. By fitting the extended model to the entire infall sample, we find that using only the merger sample can underestimate merger timescales, particularly for low mass ratios. Our model provides a valuable tool for improving semi-analytical and empirical models of galaxy formation.
\end{abstract}



\section{Introduction} \label{sec:intro}
Galaxy mergers play a fundamental role in the hierarchical structure formation paradigm, shaping the evolution of galaxies over cosmic time \citep{1972ApJ...178..623T,1978MNRAS.183..341W,1991ApJ...379...52W,1993MNRAS.262..627L}. According to the \( \Lambda \)CDM model, galaxies grow through a combination of in-situ star formation and mergers. \citep{2010gfe..book.....M,2016MNRAS.458.2371R}, making the study of mergers essential for understanding galaxy evolution. The timescale over which a merging satellite galaxy coalesces with its host influences various astrophysical processes, including star formation, black hole growth, and morphological transformation \citep{2006ApJS..163....1H}. Therefore, merger timescales are essential for semi-analytical models of galaxy formation, as it controls the merger of satellite thus determine the satellite aboundance and also the processes mentioned above \citep{1991ApJ...379...52W,1993MNRAS.264..201K,2000MNRAS.319..168C,2007MNRAS.375....2D,2011MNRAS.413..101G,2015MNRAS.451.2663H}. However, despite its importance, accurately predicting merger timescales remains a complex challenge due to the interplay of gravitational dynamics, baryonic physics, and environmental factors.

Early studies of galaxy merger timescales, primarily based on numerical simulations and analytical models, focused on the role of dynamical friction in driving satellite galaxies toward their host centers. \citet{1943ApJ....97..255C} demonstrated that the influence of random encounters on a object moving through a homogeneous and isotropic distribution of lighter objects can be effectively approximated by a dynamical friction force:
\begin{equation}
\frac{d}{dt} \mathbf{v}_{\rm sat} = -4\pi G^2 \ln(\Lambda) M_{\rm sat} \rho_{\rm host}(< v_{\rm sat}) \frac{\mathbf{v}_{\rm sat}}{v_{\rm sat}^3},
\end{equation}
where $\mathbf{v}_{\rm{sat}}$ and $M_{\rm{sat}}$ are the velocity and mass of the object, $\rho_{\rm host}(< v_{\rm sat})$ is the density of background particles with velocities less than $v_{\rm{sat}}$, and $\ln\Lambda$ is the standard Coulomb logarithm, which quantifies the ratio between the maximum and minimum impact parameters that contribute significantly to effective encounters \citep{1943ApJ....97..255C,1976MNRAS.174...19W,1987gady.book.....B}. Although Chandrasekhar's formula relies on idealized approximations and requires knowledge of $\ln\Lambda$ and the distribution function of lighter objects, it is physically well-motivated and serves as a foundational framework for understanding dynamical friction acting on satellite galaxies.

Assuming a singular isothermal sphere (SIS) background density with a Maxwellian velocity distribution, \citet{1987gady.book.....B} demonstrated that for a nearly circular orbit, the merger timescale is given by  
\begin{equation}  
\frac{T_{\rm{merger}}}{T_{\rm{dyn}}} = \frac{1.17}{\ln\Lambda} \frac{M_{\rm{host}}}{M_{\rm{sat}}}\,,  
\end{equation}  
For a satellite beginning its infall at the virial radius, \( M_{\rm{host}} \) and \( r_{\rm{host}} \) correspond to the host halo's virial mass \( M_{\rm{vir}} \) and virial radius \( r_{\rm{vir}} \), respectively. $M_{\rm{sat}}$ is the subhalo mass. The dynamical time is defined as \( T_{\rm{dyn}} = \sqrt{r_{\rm{host}}^3 / (G M_{\rm{host}})} \). \citet{1993MNRAS.262..627L} extended this result to arbitrary orbits using the orbit-averaged approximation and derived  
\begin{equation}  
\frac{T_{\rm{merger}}}{T_{\rm{dyn}}} = \frac{1.17f(\epsilon)}{\ln\Lambda} \left[\frac{r_c(E)}{r_{\rm{host}}}\right]^2 \frac{M_{\rm{host}}}{M_{\rm{sat}}}\,.  
\end{equation}  
The parameter \( \epsilon = j_{\rm{infall}}/j_c(E) \) represents the orbit's circularity, where \( r_c(E) \) and \( j_c(E) \) are the radius and specific angular momentum of a circular orbit with the same energy \( E \). Through numerical integration of the orbit-averaged equations, \citet{1993MNRAS.262..627L} found that \( f(\epsilon) \) can be accurately approximated by \( f(\epsilon) \approx \epsilon^{0.78} \), with an accuracy better than 3 percent.

However, in reality, the situation is more complex, and the assumptions used above may not always hold. For example, tidal stripping can gradually reduce the satellite's mass \citep{2001ApJ...559..716T,2022MNRAS.510.2900D,2024PhRvD.110b3019D}, an effect not accounted for in the previous calculations. Additionally, factors such as the non-spherical density distribution of host halos, the presence of substructures, the evolution of the host halo itself, and baryonic effects further complicate the merger process and are not considered in these simplified models. Therefore, to better formulate merger timescales using only initial conditions for easier application, many studies have attempted to derive more generalized fitting formulas based on N-body or hydrodynamic simulations. These efforts aim to provide a more realistic description of galaxy merger timescales \citep{1995MNRAS.275...56N, 1999ApJ...525..720C, 1999MNRAS.304..254V, 2003MNRAS.341..434T, 2008MNRAS.383...93B, 2008ApJ...675.1095J, 2010MNRAS.403.1072W, 2017MNRAS.472.1392S, 2021MNRAS.501.2810P}. The generalized fitting formulas are typically expressed as  
\begin{equation}  
\frac{T_{\rm{merger}}}{T_{\rm{dyn}}} = A\frac{f(\epsilon)}{\ln\Lambda} \left[\frac{r_c(E)}{r_{\rm{host}}}\right]^\gamma \left[\frac{M_{\rm{host}}}{M_{\rm{sat}}}\right]^\beta\,,  
\end{equation}  
where a common choice for the Coulomb logarithm is \(\Lambda = (1 + M_{\rm{host}}/M_{\rm{sat}})\). For example, using idealized N-body simulations, \citet{2008MNRAS.383...93B} found that \( f(\epsilon) = e^{1.9\epsilon} \), with \(\beta = 1.3\) and \(\gamma = 1.0\). However, results from cosmological hydrodynamic simulations by \citet{2008ApJ...675.1095J} indicate a much weaker dependence of merger timescales on initial orbits, with \( f(\epsilon) = 0.94\epsilon^{0.6} + 0.6 \), \(\beta = 1.0\), and \(\gamma = 0.5\). Furthermore, it is worth noting that \citet{2021MNRAS.501.2810P} argued that, in cosmological N-body simulations, merger timescales primarily depend on the pericentric distance of satellites from the host center, suggesting that tidal disruption or heating near pericenter may play a significant role in determining merger timescales \citep{2001ApJ...559..716T,2018MNRAS.474.3043V}.

The variation in results across different studies may suggest a fundamental limitation in the assumption that merger timescales are universally applicable to all host halos and satellites. This assumption implies that a single formula can accurately predict merger timescales based on initial conditions. However, discrepancies observed in various simulations indicate that additional factors may significantly influence these timescales. While the spherical density profile of dark matter halos exhibits a degree of universality \citep{1997ApJ...490..493N, 2000ApJ...529L..69J}, other factors that may influence merger timescales—such as the triaxial density profile \citep{2002ApJ...574..538J}, halo formation history \citep{2002ApJ...568...52W, 2003MNRAS.339...12Z, 2009ApJ...707..354Z}, and galaxy properties \citep{2010gfe..book.....M}—depend on halo mass and the surrounding environment. 

Therefore, in this paper, we investigate whether merger timescales are truly universal and, if not, seek to provide a more accurate description. To achieve this, we construct our merger samples using the \texttt{IllustrisTNG} magnetohydrodynamic cosmological simulations. Our findings reveal that the dependence of galaxy merger timescales on initial conditions varies with host halo mass, suggesting that the discrepancies observed in previous studies may stem from differences in the systems analyzed. We present a fitting formula that accounts for the dependence on host halo mass, enabling more accurate predictions of merger timescales. Moreover, we find that merger timescales differ between dark matter-only (DMO) and baryonic runs. This difference also varies with host halo mass. Finally, we demonstrate that our fitting formula can be extended to predict merger timescales for events initiated at any radius within the host halo. By incorporating the entire infall sample into the extended model, rather than just the merger sample, we also assess the impact of sample selection effects.

We introduce the simulation data and merger samples in Section \ref{sec:sim}. In Section \ref{sec:pre}, we develop a halo mass-dependent model for merger timescales. The comparison between merger timescales in DMO and baryonic runs is presented in Section \ref{sec:dmo}. In Section \ref{sec:extend}, we extend our model to arbitrary infall radii and study the sample selection effects. Finally, we provide a brief summary in Section \ref{sec:con}. 

\section{Simulation data}\label{sec:sim}
In this section, we provide a brief overview of the \texttt{IllustrisTNG} simulations. We also describe the methodology for constructing merger samples and deriving merger parameters. 

\subsection{IllustrisTNG}
The \texttt{IllustrisTNG} simulations are a suite of magnetohydrodynamic cosmological simulations \citep{2018MNRAS.480.5113M, 2018MNRAS.475..624N, 2018MNRAS.477.1206N, 2018MNRAS.475..676S, 2018MNRAS.475..648P, 2019ComAC...6....2N}. These simulations are conducted using the moving-mesh code {\texttt{AREPO}} \citep{2010MNRAS.401..791S} and incorporate a range of baryonic processes modeled as subgrid physics \citep{2017MNRAS.465.3291W, 2018MNRAS.473.4077P}. The cosmological parameters adopted in the simulations align with the Planck15 results \citep{2016A&A...594A..13P}: $\Omega_m = 0.3089$, $\Omega_{\Lambda} = 0.6911$, $\Omega_b = 0.0486$, $h = 0.6774$, $\sigma_8 = 0.8159$, and $n_s = 0.9667$. Among the suite of simulations, we utilize the \texttt{TNG100-1} and \texttt{TNG300-1} runs for this study. The \texttt{TNG100-1} simulation features a box size of $75h^{-1}{\rm{Mpc}}$, with $1820^{3}$ dark matter particles and $1820^{3}$ gas cells. This corresponds to mass resolutions of $m_{\rm{DM}} = 5.05 \times 10^{6} h^{-1} M_{\odot}$ for dark matter and average $m_{\rm{gas}} = 9.44 \times 10^{5} h^{-1} M_{\odot}$ for gas cells. The \texttt{TNG300-1} simulation has a larger box size of $205h^{-1}{\rm{Mpc}}$, containing $2500^{3}$ dark matter particles and $2500^{3}$ gas cells, with mass resolutions of $m_{\rm{DM}} = 3.98 \times 10^{7} h^{-1} M_{\odot}$ for dark matter and average $m_{\rm{gas}} = 7.44 \times 10^{6} h^{-1} M_{\odot}$ for gas cells. 

Dark matter halos in the simulation were identified using the Friends-of-Friends method, while subhalos were cataloged using the {\texttt{SUBFIND}} algorithm \citep{2001MNRAS.328..726S}. Subhalos across different snapshots were linked by merger trees constructed with the \texttt{SUBLINK} algorithm \citep{2015MNRAS.449...49R}, which traces their formation history by assigning a unique descendant (if one exists) from the subsequent snapshot to each subhalo based on a scoring system that considers shared particles.

\begin{figure}
    \centering
    \includegraphics[width=1\linewidth]{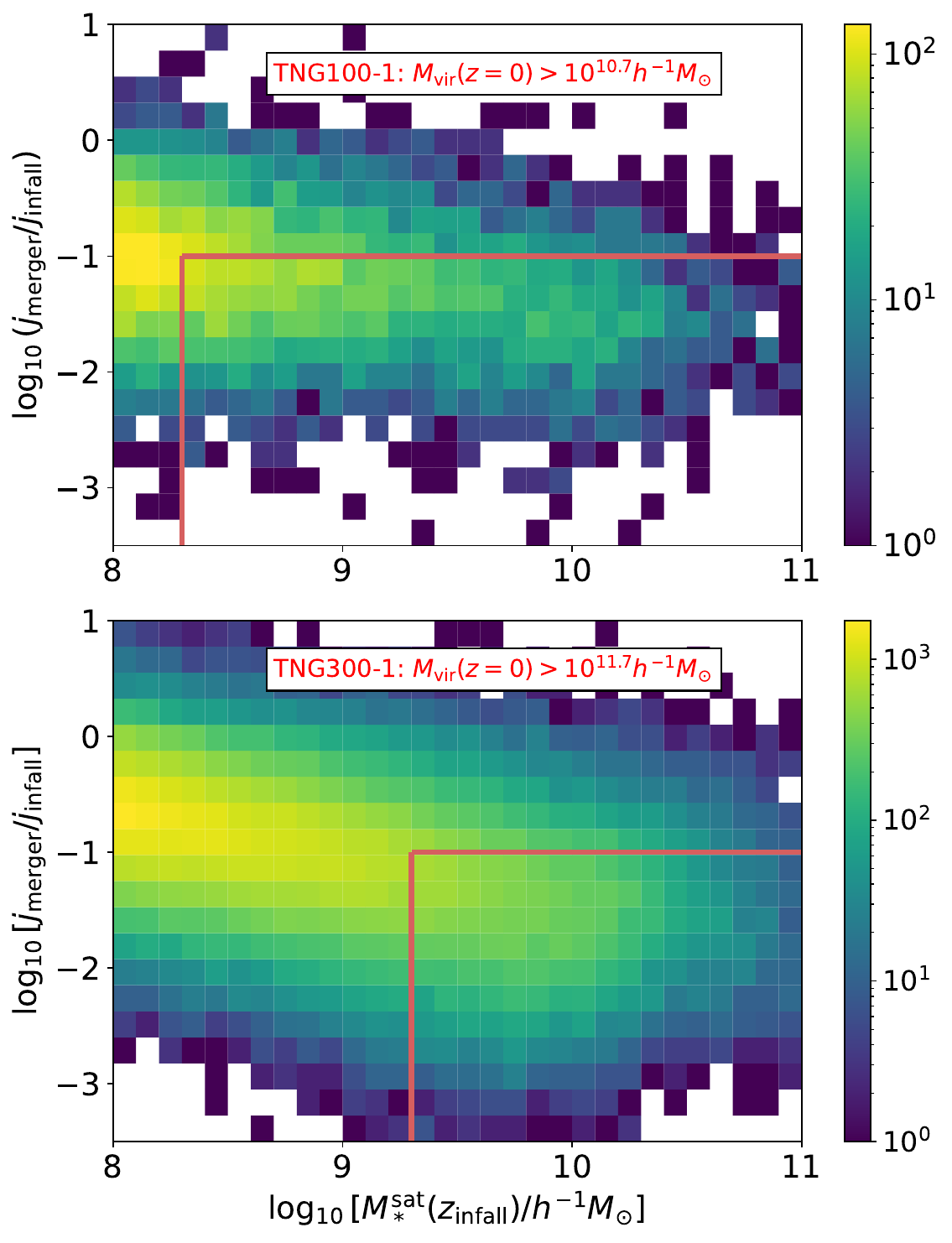}
    \caption{Joint distribution of the infall stellar mass of satellite galaxies $M_{*}^{\rm{sat}}(z_{\rm infall})$ and the remaining fraction of specific angular momentum before the merger $j_{\rm{merger}}/j_{\rm{infall}}$. Results from \texttt{TNG100-1} and \texttt{TNG300-1} are displayed in the top and bottom panels, respectively, for systems with $z_{\rm{infall}} < 4$. Red lines indicate the selection criteria used to construct our merger sample.}
    \label{fig:Msat_J}
\end{figure}

\begin{figure*}
    \centering
    \includegraphics[width=1\textwidth]{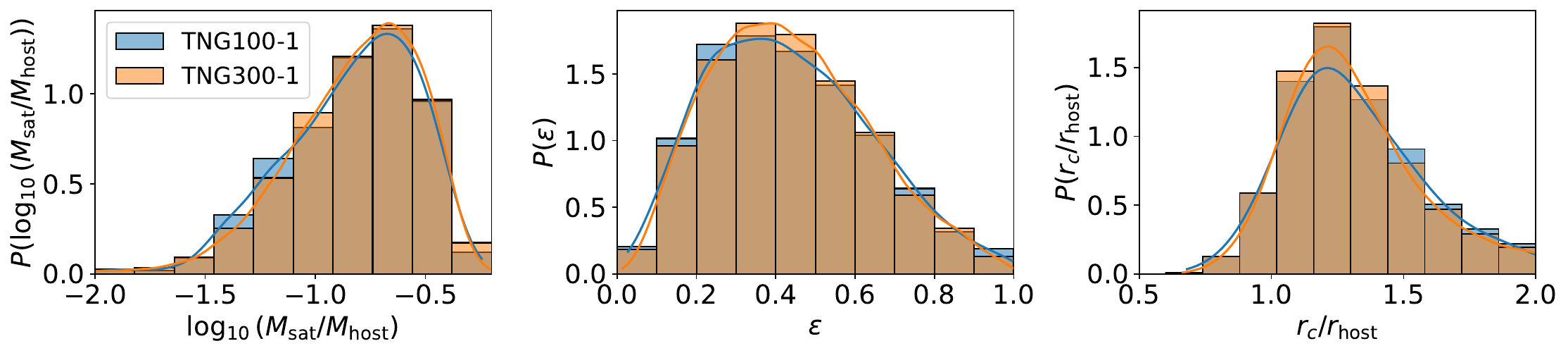}
    \caption{Distributions of $\log_{10}(M_{\rm{sat}}/M_{\rm{host}})$, $\epsilon$, and $r_c/r_{\rm{host}}$ at the infall time for our merger samples in \texttt{TNG100-1} (blue) and \texttt{TNG300-1} (orange). The lines represent kernel density estimate (KDE) distributions.}
    \label{fig:ini_dis}
\end{figure*}
\subsection{Merger samples}
We use the DM-based \texttt{SUBLINK} merger tree to study the merger timescales of galaxies. We also tested the star- and gas-based merger trees and found minimal differences in the results. We define infall time as the moment when the center of a subhalo crosses the virial radius $r_{\rm{vir}}$ of its host halo. The virial radius, $r_{\rm{vir}}$, is the radius within which the mean mass density equals $\Delta_c$ \citep{1998ApJ...495...80B} times the critical density of the universe. In the \texttt{SUBLINK} merger tree, a merger is defined as the event in which two subhalos share the same descendant. For simplicity, we consider only mergers between a satellite subhalo and a central subhalo. Specifically, we include events where the descendant of the satellite subhalo lies in the main progenitor branch of the central subhalo at $z = 0$. 

To ensure that both central and satellite galaxies are well resolved, we select host halos with virial masses $M_{\rm{vir}}(z=0) > 10^{10.7}h^{-1}M_{\odot}$ for \texttt{TNG100-1} and $M_{\rm{vir}}(z=0) > 10^{11.7}h^{-1}M_{\odot}$ for \texttt{TNG300-1} at $z = 0$, corresponding to more than 10,000 DM particles. Identifying satellite galaxies that have merged into central galaxies is nontrivial because of the challenges in robustly defining mergers in simulations. A merger, as defined by the \texttt{SUBLINK} merger tree, can correspond to one of three scenarios: a real merger, the actual disruption of a satellite \citep{2025ApJ...981..108H}, or the artificial disruption of a satellite \citep{2018MNRAS.475.4066V}. A real merger is typically defined as the point at which a large fraction of a satellite’s particles become mixed and indistinguishable from those of the central galaxy in phase space. In contrast, actual disruption refers to the physical disintegration of a satellite before its material merges with the central galaxy. The observed stellar streams in the Milky Way are remnants of such disruptions \citep{2018MNRAS.481.3442M}. Artificial disruption, on the other hand, occurs when a satellite is entirely disrupted due to limited mass and force resolution; such satellites would remain intact in higher-resolution simulations. Distinguishing among these scenarios is essential for accurately studying galaxy merger timescales, since including artificial disruption events can significantly shorten the measured durations. Since distinguishing between real and artificial disruption is challenging, we first focus on genuine merger events. Results that include disruption cases will be discussed in Section~\ref{sec:5.2}. Therefore, following \cite{2008MNRAS.383...93B}, we distinguish mergers from satellite disruptions using not only stellar mass or particle count, but also the final specific angular momentum, $j_{\rm{merger}}$, which, as shown by \cite{2008MNRAS.383...93B}, is a better indicator than $r_{\rm{merger}}$. In Figure~\ref{fig:Msat_J}, we show the joint distribution of the infall stellar mass of satellite galaxies $M_{*}^{\rm{sat}}(z_{\rm infall})$ and the ratio $j_{\rm{merger}}/j_{\rm{infall}}$ of their specific angular momentum before the merger to that at infall. The stellar mass of galaxies is defined as the total stellar mass enclosed within twice the stellar half-mass radius. We show the results for both \texttt{TNG100-1} and \texttt{TNG300-1} for all systems with $z_{\rm{infall}}<4$. It is evident that $j_{\rm{merger}}/j_{\rm{infall}}$ decreases with increasing stellar mass. Moreover, at a fixed stellar mass, \texttt{TNG100-1} exhibits, on average, lower $j_{\rm{merger}}/j_{\rm{infall}}$ values compared to \texttt{TNG300-1}. This suggests that galaxies with fewer particles are more susceptible to disruption, consistent with the results reported in \cite{2018MNRAS.475.4066V}. To construct the merger sample, we apply both stellar mass and angular momentum cuts, as indicated by the red lines in Figure~\ref{fig:Msat_J}. Specifically, we use an infall stellar mass cut of $M_*^{\rm sat}(z_{\rm infall}) > 10^{8.3} h^{-1} M_{\odot}$ for satellite galaxies in \texttt{TNG100-1} and $M_*^{\rm sat}(z_{\rm infall}) > 10^{9.3} h^{-1} M_{\odot}$ for those in \texttt{TNG300-1}, along with an angular momentum cut of $j_{\rm{merger}}/j_{\rm{infall}} < 0.1$. This stellar mass cut ensures that infalling subhalos contain more than 200 star particles and 5,000 dark matter particles, guaranteeing that the subhalos are well resolved and that baryonic effects are present. After applying the selection criteria, we identify a total of 3,792 mergers in \texttt{TNG100-1} and 22,635 mergers in \texttt{TNG300-1}.

\subsection{Merger timescales and orbit parameters}
Merger timescales refer to the time between a satellite’s infall into the host halo and its eventual disruption or merger with the central galaxy. The goal of this paper is to investigate the prediction of merger timescales for satellite galaxies based on their initial conditions at the time of infall. To achieve this, we calculate the infall and merger times, along with the orbital parameters and mass ratios at the infall time, for these satellite galaxies. 

Since the simulations are stored in discrete snapshots, determining the exact infall and merger times requires approximations. For the infall time, we use the positions, velocities, and times of satellite galaxies from the snapshots immediately before and after infall. Assuming constant acceleration in the radial direction, we solve for the precise time at which the satellite galaxies {\it first} cross \( r_{\rm{vir}} \) of their host halos. For the merger time, following \cite{2008ApJ...675.1095J}, we adopt the midpoint between the snapshots immediately before and after the merger as the merger time. The dense sampling provided by the large number of snapshots results in a time interval of $0.1-0.2\,\rm{Gyr}$ between snapshots, which only significantly affects the rare, very fast merger events.

\begin{figure*}
    \centering
    \includegraphics[width=1\textwidth]{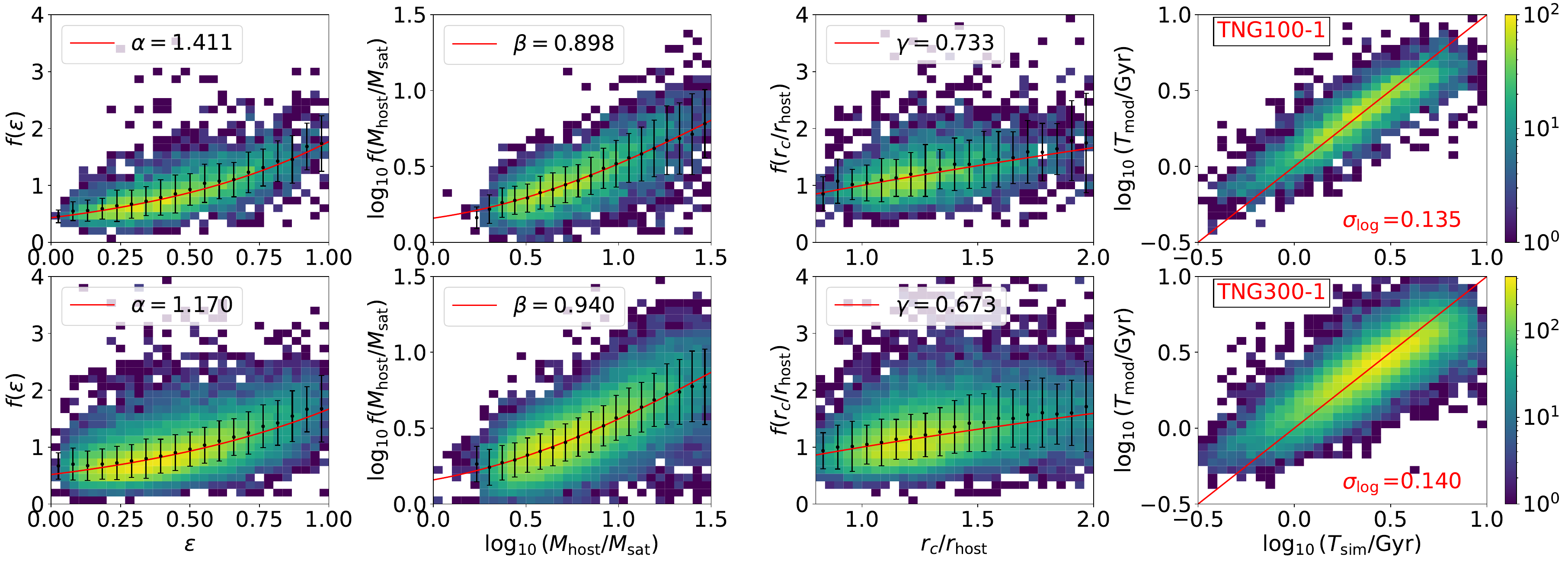}
    \caption{Comparison of the merger timescales from \texttt{TNG100-1} (top) and \texttt{TNG300-1} (bottom) with the best-fit models. The first three columns compare each component: $Ae^{\alpha\epsilon}$, $(M_{\rm{host}}/M_{\rm{sat}})^{\beta}/\ln{(1+M_{\rm{host}}/M_{\rm{sat}})}$, and $(r_c/r_{\rm{host}})^{\gamma}$. In each case, the simulation data represents the merger timescales normalized by the best-fit models excluding the respective component. The last column compares the overall merger timescales between the simulation and the models. The figures are color-coded by the number of merger events.}
    \label{fig:fit_full}
\end{figure*}

For the initial conditions, all quantities are calculated at the snapshot immediately preceding infall. Three sets of parameters are typically required to determine the merger timescale: the mass ratio $M_{\rm{sat}}/M_{\rm{host}}$, the orbit parameters $\Theta$, and the dynamical time $T_{\rm{dyn}}$. For the mass ratio, we define $M_{\rm{host}}$ as the virial mass $M_{\rm{vir}}$ of the host halo and $M_{\rm{sat}}$ as the bound subhalo mass of the satellite. The dynamical time $T_{\rm{dyn}}$ is defined as $T_{\rm{dyn}} = \sqrt{r_{\rm{host}}^3 / (G M_{\rm{host}})}$, where $r_{\rm{host}}$ is the virial radius $r_{\rm{vir}}$ of the host halos. For a spherical potential, the orbit of a satellite can be fully determined by its energy and angular momentum $(E,\ L)$. In practice, two dimensionless quantities are commonly used: $\epsilon$ and $r_c(E)/r_{\rm{host}}$. Here, $\epsilon = j_{\rm{infall}}/j_c(E)$ represents the circularity of the orbit, while $r_c(E)$ and $j_c(E)$ denote the radius and specific angular momentum, respectively, of a circular orbit with the same energy $E$. As in many studies \citep{1993MNRAS.262..627L, 2008ApJ...675.1095J, 2015MNRAS.448.1674J}, we adopt a SIS potential for halos to calculate the initial orbit parameters, as it is more convenient than the Navarro–Frenk–White \citep[NFW;][]{1997ApJ...490..493N} potential. The differences in orbit parameters between the SIS and NFW potentials have been explored in \citet[][see Figure~11]{2015MNRAS.448.1674J}. While the NFW potential may be more realistic, we find that the parameters derived from the SIS potential already provide accurate predictions for merger timescales, as indicated by $\sigma_{\log}$ below. Given its simplicity and effectiveness, we retain the use of the SIS potential in our analysis.

In Figure~\ref{fig:ini_dis}, we show the distributions of the initial conditions $M_{\rm{sat}}/M_{\rm{host}}$, $\epsilon$, and $r_c(E)/r_{\rm{host}}$ for our merger samples. Despite the differences in the selection criteria for $M_{*}^{\rm{sat}}$ and $M_{\rm{vir}}$ between TNG100-1 and TNG300-1, their merger samples exhibit very similar initial distributions. Similar to \cite{2008ApJ...675.1095J}, most mergers in our sample have a mass ratio larger than 0.03, as smaller satellites cannot merge within a Hubble time. Additionally, we find that our merger sample exhibits smaller $\epsilon$ values compared to the complete infall sample studied in \cite{2015MNRAS.448.1674J}, while the $r_c(E)/r_{\rm{host}}$ distribution is quite similar. This arises because radial orbits have shorter merger timescales and are more likely to be included in our merger sample. The impact of these selection effects on our results will be examined in Section \ref{sec:5.2} using the full infall sample rather than only merger sample.

\section{Predicting Merger timescales}\label{sec:pre}
In this section, we present a model for predicting the merger timescales of satellite galaxies based on their initial conditions at infall. Additionally, we test the model's universality and examine its dependence on host halo mass.

\subsection{Modeling the entire merger samples}
As in previous studies \citep{2008ApJ...675.1095J,2017MNRAS.472.1392S,2021MNRAS.501.2810P}, we begin by attempting to construct a universal model for the entire merger sample. We adopt the exponential form of \( f(\epsilon) \) from \cite{2008MNRAS.383...93B}, as it provides a better fit to our results than alternative forms such as power laws. The merger timescale model we adopt is  
\begin{equation}
    \frac{T_{\rm{merger}}}{T_{\rm{dyn}}} = A e^{\alpha\epsilon} \frac{(M_{\rm{host}}/M_{\rm{sat}})^{\beta}}{\ln(1+M_{\rm{host}}/M_{\rm{sat}})} \left[\frac{r_c(E)}{r_{\rm{host}}}\right]^{\gamma}\,. \label{eq:time_scale}
\end{equation}
Here, \( \alpha \), \( \beta \), and \( \gamma \) control the dependence of merger timescales on orbital circularity, mass ratio, and orbital energy, respectively, while \( A \) serves as a normalization factor.  

We fit this model to our merger samples from \texttt{TNG100-1} and \texttt{TNG300-1} separately using the following expression for chi-squared:
\begin{equation}
    \chi^2 = \frac{\sum_{i=1}^N[\log_{10}(T_{\rm{sim,i}}) - \log_{10}(T_{\rm{mod,i}})]^2}{N-p}\,,
\end{equation}
where $N$ is the number of mergers in the sample, $p$ is the number of free parameters, $T_{\rm{sim}}$ represents the merger timescales calculated from simulations, and $T_{\rm{mod}}$ denotes the timescales predicted by the model. The \( \chi^2 \) is minimized using the Levenberg–Marquardt algorithm. We obtain the following best-fit parameters for \texttt{TNG100-1}:  
\begin{align}
    A &= 0.431 \pm 0.009, \quad \alpha = 1.411 \pm 0.025 \notag \\
    \beta &= 0.898 \pm 0.008, \quad \gamma = 0.733 \pm 0.020 \notag
\end{align}  
with a logarithmic scatter of $\sigma_{\rm{log}} = 0.135$. For \texttt{TNG300-1}, the best-fit parameters are  
\begin{align}
    A &= 0.518 \pm 0.004, \quad \alpha = 1.170 \pm 0.011 \notag \\
    \beta &= 0.940 \pm 0.004, \quad \gamma = 0.673 \pm 0.009 \notag
\end{align}  
with a logarithmic scatter of $\sigma_{\rm{log}} = 0.140$.

In Figure~\ref{fig:fit_full}, we compare the merger timescales from simulations and best-fit models. Each component in Equation~\ref{eq:time_scale} is examined separately, with the simulation data showing merger timescales normalized by the best-fit models that exclude the corresponding component. We find that the dependence of merger timescales on all components—$\epsilon$, $M_{\rm{sat}}/M_{\rm{host}}$, and $r_c/r_{\rm{host}}$—is well captured by Equation~\ref{eq:time_scale}. The overall scatter in both \texttt{TNG100-1} and \texttt{TNG300-1} is small with $\sigma_{\log}$ smaller than 0.15, indicating that our model provides accurate predictions for merger timescales. However, although our model fits the merger timescales in both \texttt{TNG100-1} and \texttt{TNG300-1} samples well, the best-fit parameters differ, particularly for $\alpha$, which describes the dependence on orbit circularity $\epsilon$. This suggests that galaxy merger timescales may not be universal and could depend on sample properties.

\begin{table*}
    \centering
    \caption{Descriptions and best-fit parameters of the four merger timescale models constructed in this study, based on different samples, starting radii, and model parameterizations. The first two rows use the merger sample starting at \( r_{\rm vir} \), from baryonic and DMO runs respectively. The third uses the merger sample with arbitrary starting radii in baryonic runs. The fourth, recommended for general use, is based on the full infall sample with arbitrary starting radii in baryonic runs.}
    \footnotesize 
    \begin{tabular}{cccccccccccccc}
    \hline
     sample& start & model&$A_{\rm{h}}$&$M_{\rm{min}}$& $\sigma_{\log M}$&$A_{\rm{min}}$&$\alpha_{\rm{h}}$& $\alpha_{\rm{min}}$&$\gamma_{\rm{h}}$&$\gamma_{\rm{min}}$& $\beta_1$& $\beta_2$ & $M_{\beta}$ \\ 
     \hline
     merger hydro& $r_{\rm{vir}}$ & Eq. \ref{eq:time_scale}, \ref{eq:mass_dependence}&0.358 & 12.447 & 0.551 & 0.756 & 0.537 & 0.575 &  0.193 & 0.405 & 0.806&&\\   
     merger DMO & $r_{\rm{vir}}$ & Eq. \ref{eq:time_scale}, \ref{eq:mass_dependence} & 0.257 & 13.228 & 0.885 & 0.987 & 0.393 & 0.208 & 0.332 & 0.038 & 0.824&&\\   
     merger hydro& any & Eq. \ref{eq:mass_dependence}, \ref{eq:time_scale_ex}, \ref{eq:factor},  \ref{eq:mass_double}, \ref{eq:Tdy} & 0.360 & 12.376 & 0.605 & 0.833 & 0.406 & 0.700 & -0.026 & 0.892 & 0.814 & 94.736 & -0.456\\
     infall hydro& any & Eq. \ref{eq:mass_dependence}, \ref{eq:time_scale_ex}, \ref{eq:factor},  \ref{eq:mass_double}, \ref{eq:Tdy} & 0.323 & 12.484 & 0.555 & 0.657 & 0.328 & 0.497 & 0.326 & 0.408 & 1.077 & 30.890 & -0.026\\
     \hline
    \end{tabular}
    \label{tab:best-fit}
\end{table*}

\begin{figure*}
    \centering
    \includegraphics[width=1\textwidth]{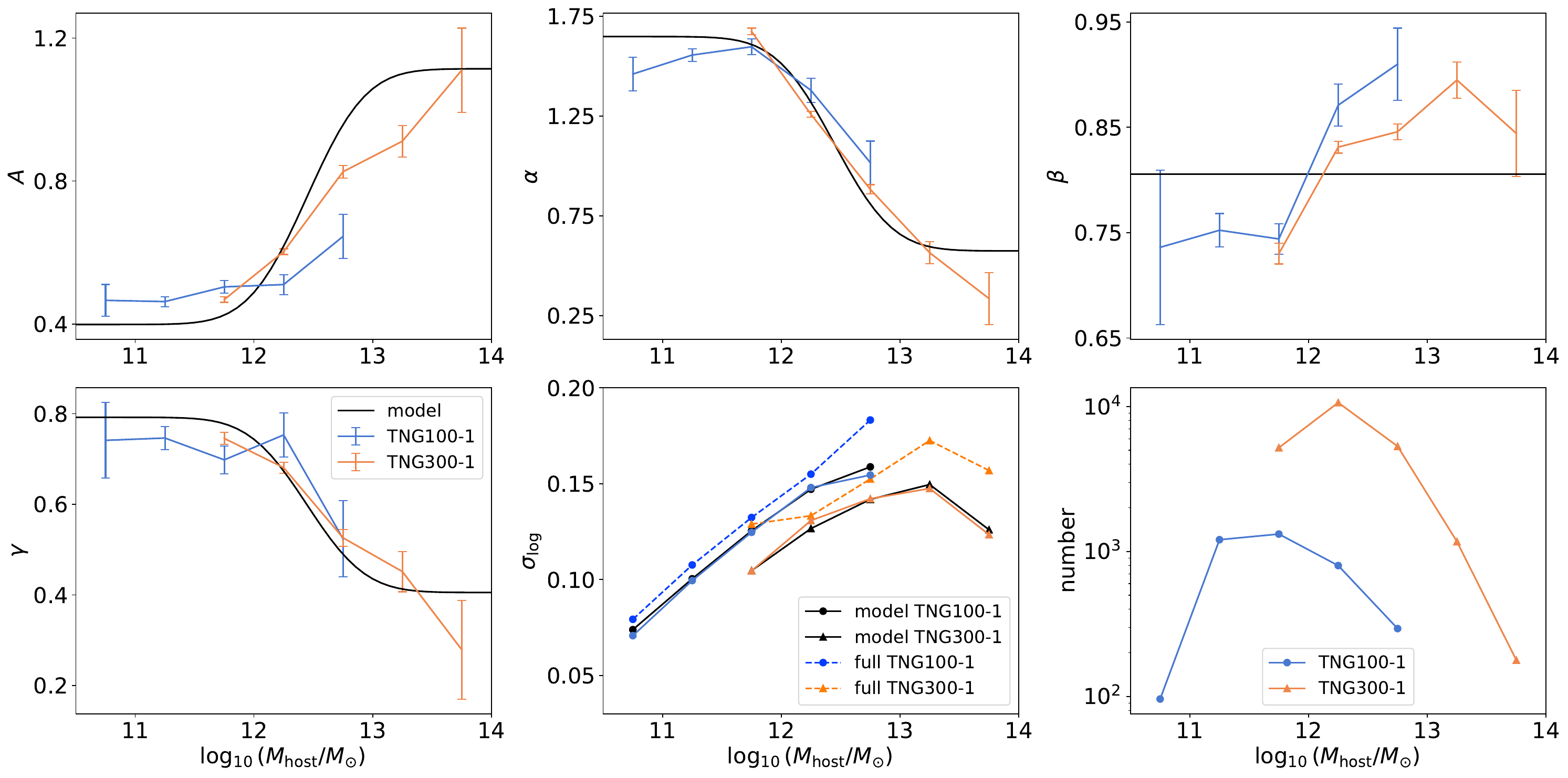}
    \caption{The dependence of galaxy merger timescale parameters on the infall host halo mass $M_{\rm{host}}$ is shown. Results from \texttt{TNG100-1} (blue) and \texttt{TNG300-1} (orange) are presented separately, with the best-fit models shown as black solid lines.}
    \label{fig:param_model}
\end{figure*}

\subsection{Dependence on infall host halo mass}
 The differences in best-fit models between \texttt{TNG100-1} and \texttt{TNG300-1}, along with inconsistencies in previous studies such as \cite{2008ApJ...675.1095J} and \cite{2008MNRAS.383...93B}, suggest that galaxy merger timescales may depend on sample properties. Therefore, we divide the merger samples from \texttt{TNG100-1} and \texttt{TNG300-1} into different infall $M_{\rm{host}}$ bins and model them separately, as host halo mass is a fundamental property that influences key factors such as halo formation history, concentration, ellipticity, and subhalo distributions.  
 
The best-fit parameters and their uncertainties are shown in Figure~\ref{fig:param_model}, while the data and best-fit models are presented in Figures \ref{fig:fit_tng100} and \ref{fig:fit_tng300}, respectively. We find that the parameters indeed depend on $M_{\rm{host}}$ and the results in \texttt{TNG100-1} and \texttt{TNG300-1} are now consistent. Specifically, the orbital parameters $\alpha$ and $\gamma$ decrease with increasing $M_{\rm{host}}$, while remaining relatively unchanged for $M_{\rm{host}} < 10^{12} M_{\odot}$. This suggests that the dependence of galaxy merger timescales on the initial orbit weakens in higher-mass host halos.  For the mass ratio dependence, $\beta$ increases slightly with $M_{\rm{host}}$, from 0.75 to approximately 0.9. Additionally, the scatter $\sigma_{\rm{log}}$ also increases with $M_{\rm{host}}$, further indicating a weaker connection between merger timescales and initial conditions in more massive host halos.

Comparing our results with previous findings from \cite{2008ApJ...675.1095J} and \cite{2008MNRAS.383...93B}, we find that for $M_{\rm{host}} < 10^{12} M_{\odot}$, the parameters $\alpha$ and $\gamma$ approach the values of 1.9 and 1.0, respectively, as reported in the idealized simulations of \cite{2008MNRAS.383...93B}. At higher host masses, our results align with the conclusion of \cite{2008ApJ...675.1095J}, which suggests that the dependence on initial orbital conditions becomes relatively weak. This may explain the discrepancies between these two studies as they are studying systems in different conditions. It is reasonable that in the  idealized simulations of \cite{2008MNRAS.383...93B} merger timescales show the strongest connection to the initial conditions. In realistic simulations, violent processes in host halos, such as major mergers \citep{2019MNRAS.487..993D,2019MNRAS.487.1008D} and interactions between satellite galaxies, can make the infall process more collisional, thereby erasing initial conditions and weakening their connection to merger timescales. For example, a sudden shift in the halo center position following a major merger can significantly weaken the connection, as it alters the reference frame used to calculate the orbital parameters.
. According to our results, these processes may occur more frequently in more massive host halos.

To characterize the dependence of merger timescales on $M_{\rm{host}}$, we parameterize $A$, $\alpha$, $\beta$, and $\gamma$ as functions of $M_{\rm{host}}$. Based on Figure~\ref{fig:param_model}, we adopt the following parametrization:
\begin{align}
    A &= A_{\rm{h}}{\rm{erf}}\left(\frac{M_{\rm{host}}-M_{\rm{min}}}{\sigma_{\log M}}\right)+A_{\rm{min}}\,,\notag\\
    \alpha &= \alpha_{\rm{h}}{\rm{erfc}}\left(\frac{M_{\rm{host}}-M_{\rm{min}}}{\sigma_{\log M}}\right)+\alpha_{\rm{min}}\,,\notag\\
     \gamma &= \gamma_{\rm{h}}{\rm{erfc}}\left(\frac{M_{\rm{host}}-M_{\rm{min}}}{\sigma_{\log M}}\right)+\gamma_{\rm{min}}\,,\label{eq:mass_dependence}
\end{align}
and a constant $\beta=\beta_1$, where $\rm{erf}$ is the error function and $\rm{erfc}$ is the complementary error function. These functional forms based on the error function are chosen to interpolate between two limiting cases: the strongest orbital dependence, as found in idealized simulation by \cite{2008MNRAS.383...93B} with $\alpha = 1.9$ and $\gamma = 1.0$, and the weakest case of no dependence, corresponding to $\alpha = 0$ and $\gamma = 0$. Therefore, the model contains 9 parameters: $ \{A_{\rm{h}}, M_{\rm{min}}, \sigma_{\log M}, A_{\rm{min}},\alpha_{\rm{h}}, \alpha_{\rm{min}},\gamma_{\rm{h}},\gamma_{\rm{min}}, \beta_1 \}$. We fit the model directly to the merger timescales of the combined \texttt{TNG100-1} and \texttt{TNG300-1} sample. The best-fit model is represented by black solid lines in Figure~\ref{fig:param_model}, with the corresponding parameters listed in the first row of Table \ref{tab:best-fit}. We find that the best-fit model generally follows the trends observed when modeling different $M_{\rm{host}}$ bins separately. Additionally, we obtain a similar $\sigma_{\rm{log}}$ as in the $M_{\rm{host}}$ bin results, indicating that our model provides a sufficiently accurate description.

\begin{figure*}
    \centering
    \includegraphics[width=1\textwidth]{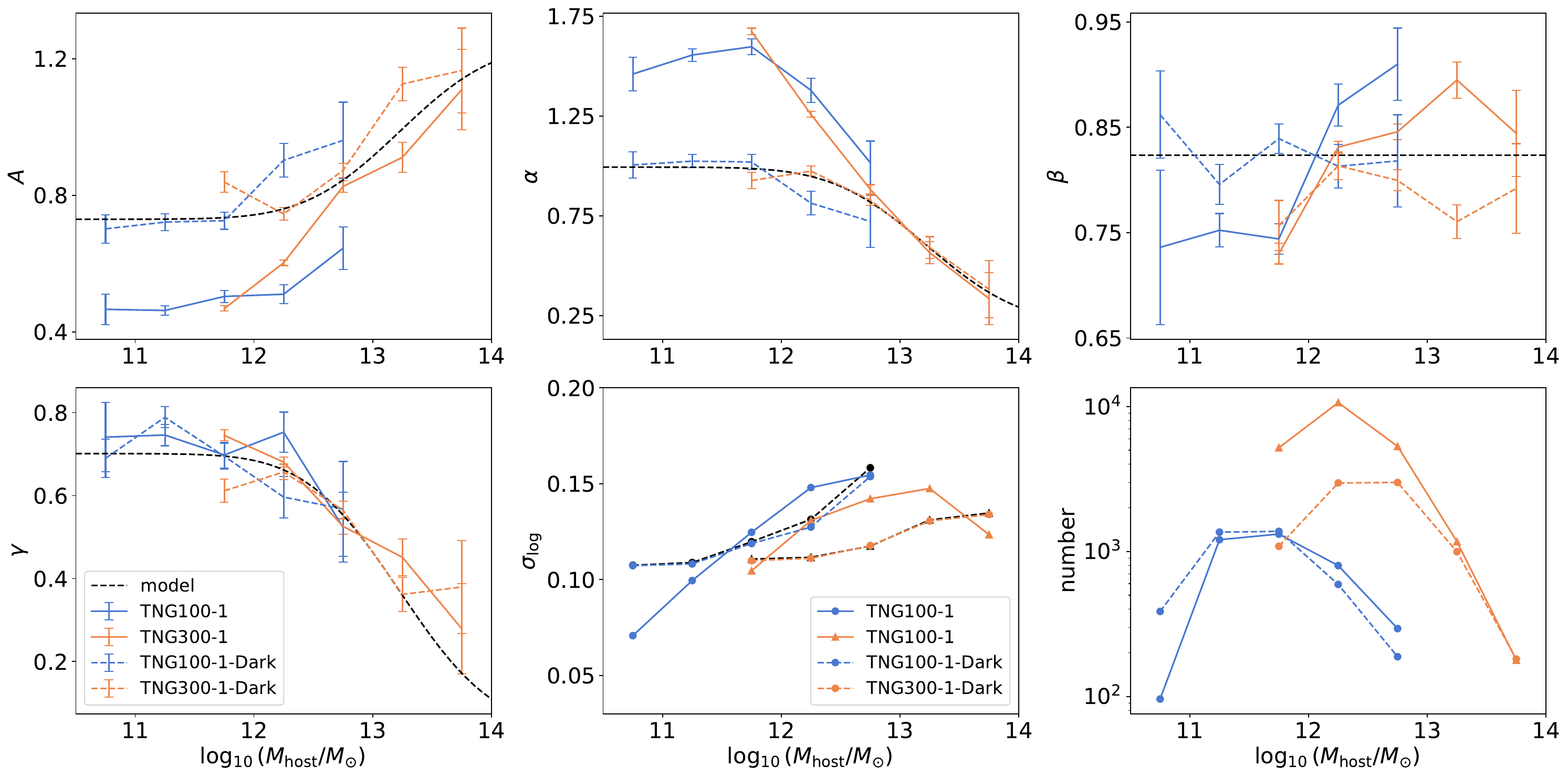}
    \caption{Comparison of merger timescale parameters between DMO (dashed lines) and hydrodynamic (solid lines) simulations in \texttt{IllustrisTNG}. The best-fit models for DMO merger samples are also shown with black dashed lines.}
    \label{fig:param_model_dark}
\end{figure*}

In Figure~\ref{fig:param_model}, we also present $\sigma_{\rm{log}}$ for different $M_{\rm{host}}$ using the parameters from the full-sample fit shown in Figure~\ref{fig:fit_full}. The $\sigma_{\rm{log}}$ from the full-sample model is higher than that from the $M_{\rm{host}}$-dependent model, confirming that incorporating $M_{\rm{host}}$ dependence improves the model's accuracy. We also note that \texttt{TNG300-1} shows a smaller \( \sigma_{\rm{log}} \) than \texttt{TNG100-1} at fixed \( M_{\rm host} \). This is because the merger samples in \texttt{TNG100-1} span a broader mass ratio range—extending to lower values—due to a lower \( M_*^{\rm sat}(z_{\rm infall}) \) cut. If merger timescales retain some dependence on \( M_{\rm sat} \), this broader range would naturally lead to a larger \( \sigma_{\rm{log}} \) in \texttt{TNG100-1}, which appears to be the case. We do not pursue a detailed investigation of the \( M_{\rm sat} \) dependence, as the current model already yields a relatively low \( \sigma_{\rm{log}} \).

\section{Comparison with Dark Matter Only Simulations}\label{sec:dmo}
In this section, we examine the differences in galaxy merger timescales between dark matter only and hydrodynamic simulations and discuss the possible underlying causes.

\begin{figure*}
    \centering
    \includegraphics[width=1\textwidth]{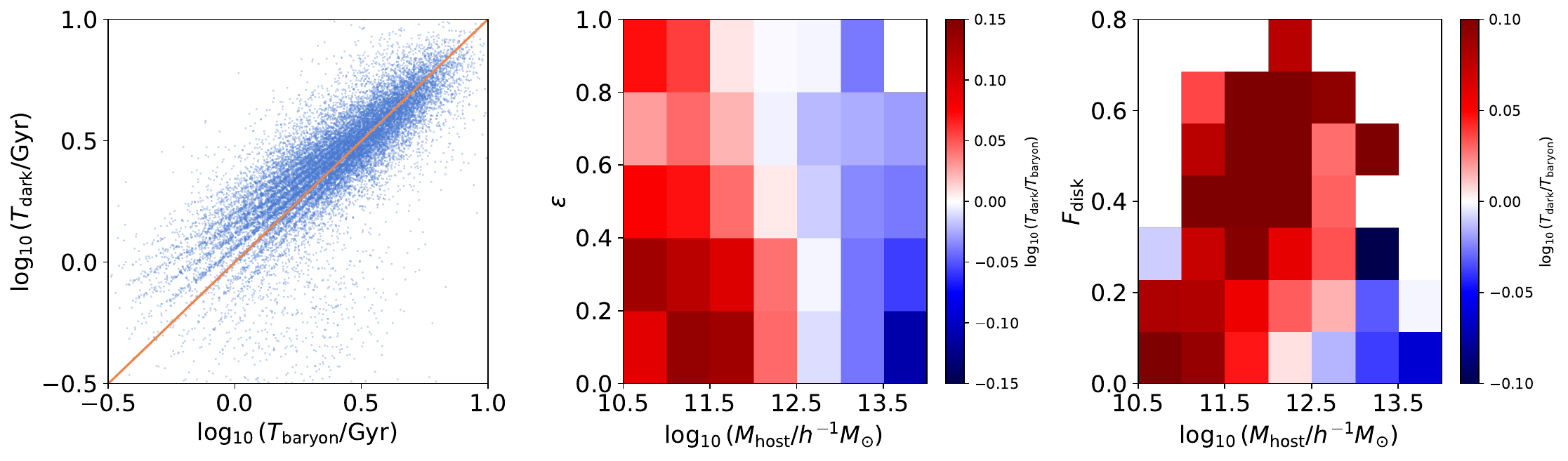}
    \caption{Comparison of merger timescales for matched events between DMO and baryonic runs. Left: Direct comparison of merger timescales. Middle: Logarithmic difference in merger timescales, \(\log_{10}(T_{\rm{dark}}/T_{\rm{baryon}})\), binned by \(M_{\rm{host}}\) and \(\epsilon\). Right: Logarithmic difference in merger timescales, \(\log_{10}(T_{\rm{dark}}/T_{\rm{baryon}})\), binned by \(M_{\rm{host}}\) and disk fraction \(F_{\rm{disk}}\).}
    \label{fig:one2one}
\end{figure*}

\begin{figure*}
    \centering
    \includegraphics[width=1\textwidth]{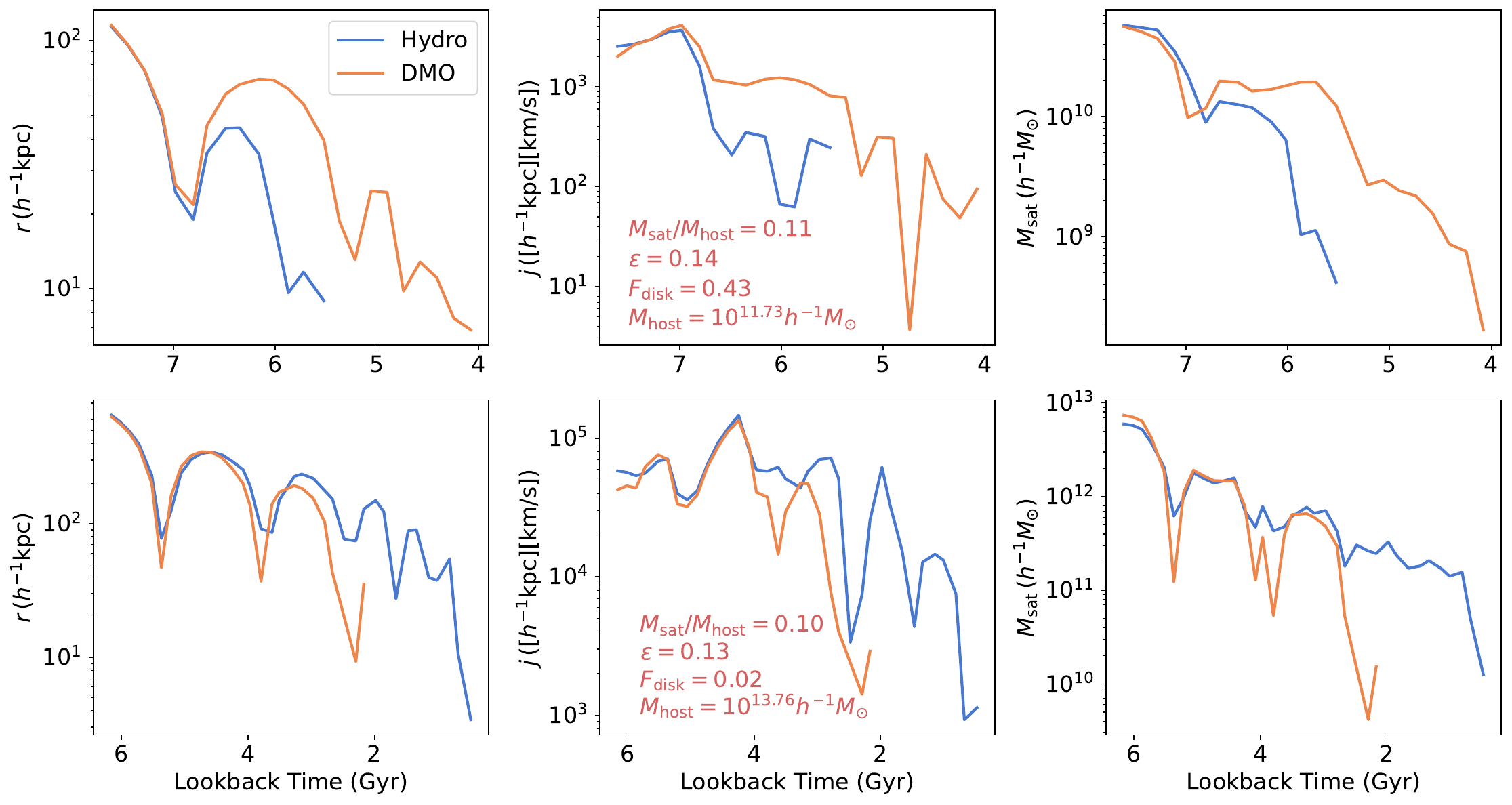}
    \caption{Comparison of the infall history of two matched events between DMO and baryonic runs. One occurs in a low-mass halo (\(M_{\rm{host}}=10^{11.73}h^{-1}M_{\odot}\)) hosting a central disk galaxy (\(F_{\rm{disk}}=0.42\)) in the baryonic run, while the other takes place in a massive halo (\(M_{\rm{host}}=10^{13.76}h^{-1}M_{\odot}\)) with a central elliptical galaxy (\(F_{\rm{disk}}=0.02\)). Both events have similar radial orbits (\(\epsilon \sim 0.13\)) and mass ratios (\(M_{\rm{sat}}/M_{\rm{host}} \sim 0.1\)). The Figure~shows the trajectories \(r\) (left), the evolution of specific angular momentum \(j\) (middle), and the bound mass of the satellite (right) for each event.}
    \label{fig:history}
\end{figure*}

\subsection{Dark matter only analogs}
To investigate the differences between DMO and hydrodynamic simulations, we use the DMO simulations \texttt{TNG100-1-Dark} and \texttt{TNG300-1-Dark}, which serve as the DMO counterparts to \texttt{TNG100-1} and \texttt{TNG300-1}. These simulations share the same cosmology, box sizes, and initial conditions. The mass resolutions of \texttt{TNG100-1-Dark} and \texttt{TNG300-1-Dark} are \( m_{\rm{DM}} = 6.00 \times 10^{6} h^{-1} M_{\odot} \) and \( 4.73 \times 10^{7} h^{-1} M_{\odot} \), respectively. \texttt{IllustrisTNG} provides matched catalogs between DMO and baryonic runs using the \texttt{SUBLINK} weighting algorithm, which is similar to the method used for tracking descendants across different snapshots within the same simulation.

\subsection{Statistical comparison}
We start with a statistical comparison between DMO and baryonic runs, calculating merger timescales and their dependence on initial conditions in the DMO simulations using the same methodology as in the baryonic simulations. To construct the merger sample, we apply a selection similar to that in Figure~\ref{fig:Msat_J}. Specifically, we select host halos with virial masses of \( M_{\rm{vir}} > 10^{10.7} h^{-1} M_{\odot} \) for \texttt{TNG100-1-Dark} and \( M_{\rm{vir}} > 10^{11.7} h^{-1} M_{\odot} \) for \texttt{TNG300-1-Dark} at \( z = 0 \). Additionally, we require subhalo bound masses at the infall snapshot to be \( M_{\rm{sat}} > 10^{10.5} h^{-1} M_{\odot} \) for \texttt{TNG100-1-Dark} and \( M_{\rm{sat}} > 10^{11.5} h^{-1} M_{\odot} \) for \texttt{TNG300-1-Dark}. The same selection criteria of \( j_{\rm{merger}}/j_{\rm{infall}} < 0.1 \) and \( z_{\rm{infall}} < 4 \) are also applied. Next, we calculate the merger timescales and initial conditions at the infall time and derive the merger timescale models for different infall host halo masses \( M_{\rm{host}} \) according to Equation~\ref{eq:time_scale}. Similar to the baryonic runs, we fit the results from the DMO merger sample using the models in Equation~\ref{eq:mass_dependence}. The best-fit parameters are provided in the second row of Table \ref{tab:best-fit}, and the corresponding best-fit models are shown in Figure~\ref{fig:param_model_dark}.

Figure~\ref{fig:param_model_dark} compares merger timescale parameters between DMO and baryonic runs. We find that \(\alpha\) is lower in DMO runs, with the difference increasing as \(M_{\rm{host}}\) decreases, while \(\beta\) and \(\gamma\) remain largely similar. As a result of the variation in $\alpha$, the normalization parameter $A$ also differs. To investigate the cause of this difference, we compare the best-fit models across different \(M_{\rm{host}}\) bins for \texttt{TNG100-1-Dark} and \texttt{TNG100-1} in Figure~\ref{fig:fit_tng100_dark}, as well as for \texttt{TNG300-1-Dark} and \texttt{TNG300-1} in Figure~\ref{fig:fit_tng300_dark}. We find that, in low-mass host halos, merger timescales in DMO simulations are generally longer than those in hydrodynamic simulations for highly radial orbits with smaller \(\epsilon\). To determine what cause of the difference, we further conduct a detailed one-to-one comparison of merger events in DMO and baryonic runs.

\subsection{One-to-one comparison}
Since \texttt{IllustrisTNG} provides matched catalogs between DMO and baryonic runs, it offers a more effective way to study the factors influencing merger timescales through a one-to-one comparison. To match the events, we first identify the corresponding host halos in the DMO runs for those in the baryonic runs at \( z=0 \). Then, for each merged satellite in the baryonic runs, we search for its DMO counterpart among the progenitors of the matched host halos at the satellite's infall snapshot. Using this approach, we identify 3,103 matched pairs out of 3,792 in \texttt{TNG100-1-Dark} and 18,871 out of 22,635 in \texttt{TNG300-1-Dark}.

In the left panel of Figure~\ref{fig:one2one}, we compare the merger timescales \(T_{\rm{baryon}}\) from the baryonic runs with \(T_{\rm{dark}}\) from the corresponding matched events in the DMO runs. The results exhibit a quantized pattern because merger times are determined using the midpoint between the snapshots immediately before and after the merger. The differences between baryonic and DMO runs appear larger for short-duration mergers, while they remain fairly consistent for long-duration mergers. This is consistent with the results of the statistical comparison, as short-duration mergers typically occur in more radial orbits with smaller \(\epsilon\). To better quantify the dependence of the merger timescale difference on \(M_{\rm{host}}\) and \(\epsilon\), the middle panel presents the logarithmic difference \(\log_{10}(T_{\rm{dark}}/T_{\rm{baryon}})\) across different \(M_{\rm{host}}\) and \(\epsilon\) bins. For each event, we choose \(M_{\rm{host}}\) and \(\epsilon\) from the baryonic runs. We find that events with smaller \(\epsilon\) indeed exhibit larger differences. However, interestingly, events with large and small \(M_{\rm{host}}\) exhibit different trends. For \(M_{\rm{host}} < 10^{12.5} h^{-1} M_{\odot}\), merger timescales are larger in the DMO runs, whereas for \(M_{\rm{host}} > 10^{12.5} h^{-1} M_{\odot}\), they are larger in the baryonic runs. Additionally, the magnitude of this difference increases toward both smaller and larger \(M_{\rm{host}}\). 

The differing impact of baryons on either side of \(M_{\rm{host}} \sim 10^{12.5} h^{-1} M_{\odot}\) suggests that baryonic effects can influence merger timescales in multiple ways, either extending or shortening them. The dominant effect may vary depending on \(M_{\rm{host}}\), leading to the observed trend. Firstly, the presence of a disk structure in the central galaxy can more effectively disrupt infalling satellites, shortening merger timescales—particularly for those on radial orbits that pass through the central region \citep{1992ApJ...389....5T,1999MNRAS.304..254V,2003ApJ...592L..25H,2017MNRAS.471.1709G,2020MNRAS.491.1471S}. Secondly, the presence of baryons in the satellite galaxy may increase the inner density of the subhalo, making it more resistant to tidal disruption. If the tidal forces in the inner region of the host halo are strong enough to disrupt the subhalos in DMO simulations, the survival of the corresponding subhalo in baryonic simulations may lead to longer merger timescales. However, if the tidal forces are not strong enough to completely disrupt the subhalo but instead significantly reduce its mass in the DMO simulation, longer merger timescales in DMO runs are expected, as reported in \citet{2008MNRAS.383...93B}. Finally,  in halos hosting high-mass elliptical galaxies, strong feedback from the supermassive black hole reduces the matter density in the central region of the host halos \citep[][see Figure~17]{2024ApJ...973..102X}, primarily within \( r < 100 h^{-1} {\rm{kpc}} \). This reduction in density may weaken the deceleration caused by dynamical friction, as described by Chandrasekhar's formula, leading to an increase in merger timescales. 

\begin{figure*}
    \centering
    \includegraphics[width=1\textwidth]{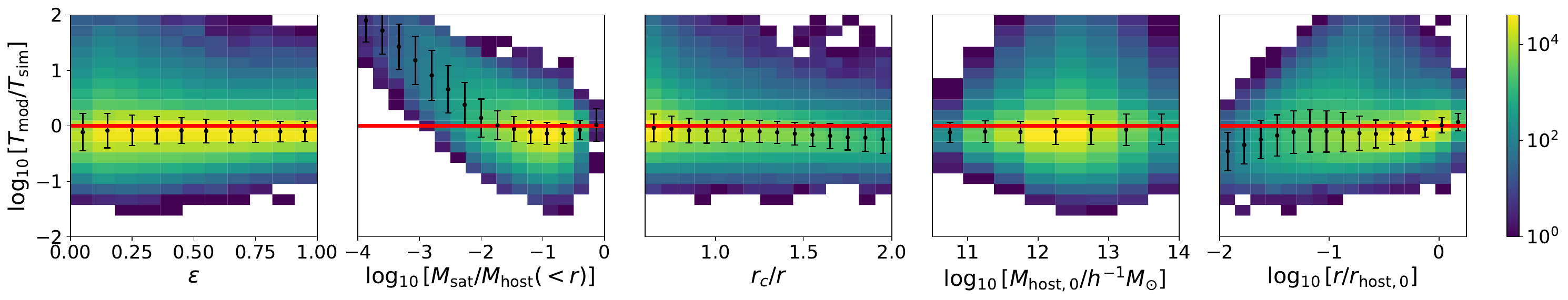}
    \caption{Comparison of merger timescales starting at any radius between simulation results and model predictions using best-fit parameters from those starting at \( r_{\rm{vir}} \). The logarithmic difference, \( \log_{10}[T_{\rm{mod}}/T_{\rm{sim}}] \), is shown as a function of orbit circularity (\(\epsilon\)), mass ratio (\(M_{\rm sat}/M_{\rm host}(<r)\)), orbit energy (\(r_c/r\)), host halo mass at the time of crossing \( r_{\rm vir} \) (\(M_{{\rm host},0}\)), and infall radius (\(r\)). The figures are color-coded by the number of arbitrary starting merger events.}
    \label{fig:extend_test}
\end{figure*}

\begin{figure}
    \centering
    \includegraphics[width=1\linewidth]{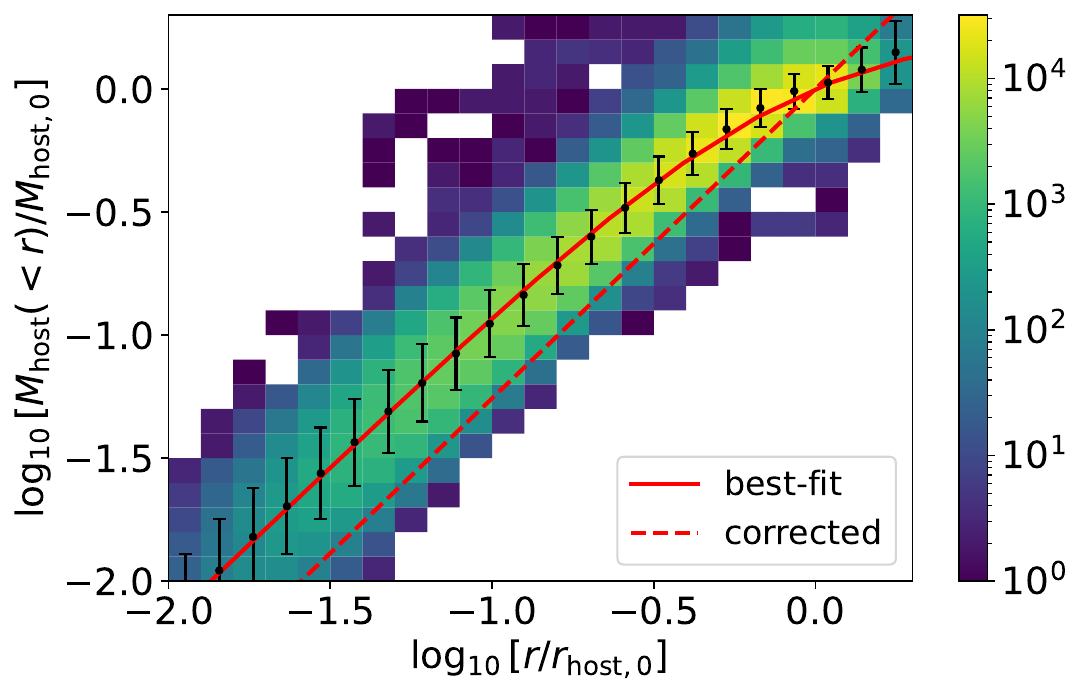}
    \caption{The enclosed mass fraction, \( M_{\rm host}(<r)/M_{\rm host,0} \), is shown as a function of the infall radius, \( r/r_{\rm host,0} \). The red solid and dashed lines represent the best-fit and corrected relation. The figures are color-coded by the number of arbitrary starting merger events.}
    \label{fig:mass_correct}
\end{figure}

\begin{figure*}
    \centering
    \includegraphics[width=1\textwidth]{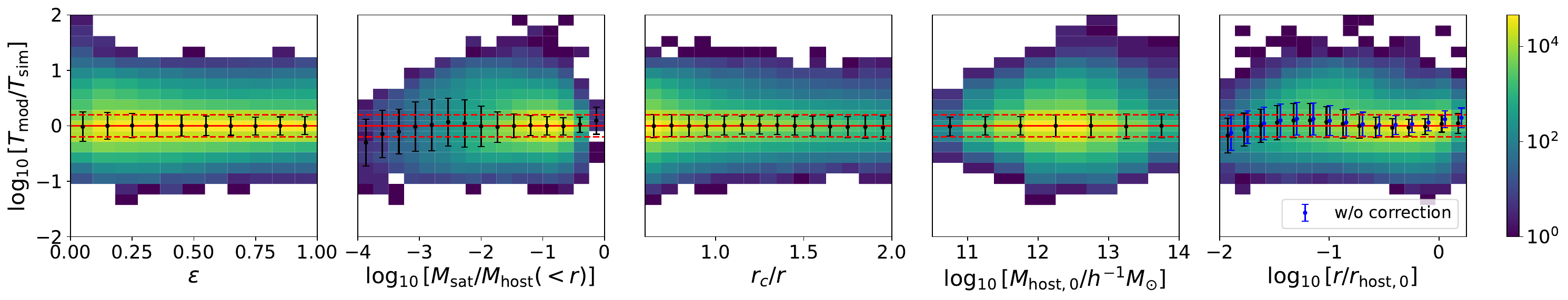}
    \caption{Similar to Figure~\ref{fig:extend_test}, but showing a comparison between simulation results and predictions from the new model, which combines Equations \ref{eq:mass_dependence}, \ref{eq:time_scale_ex}, \ref{eq:factor}, \ref{eq:mass_double}, and \ref{eq:Tdy}. For reference, the last panel also presents the best-fit results for the model without corrections from Equation~\ref{eq:Tdy} for $T_{\rm dy}$. The red dashed lines indicate scatter values of -0.2 and 0.2. The figures are color-coded by the number of arbitrary starting merger events.}
    \label{fig:extend_fit}
\end{figure*}

\begin{figure*}
    \centering
    \includegraphics[width=1\textwidth]{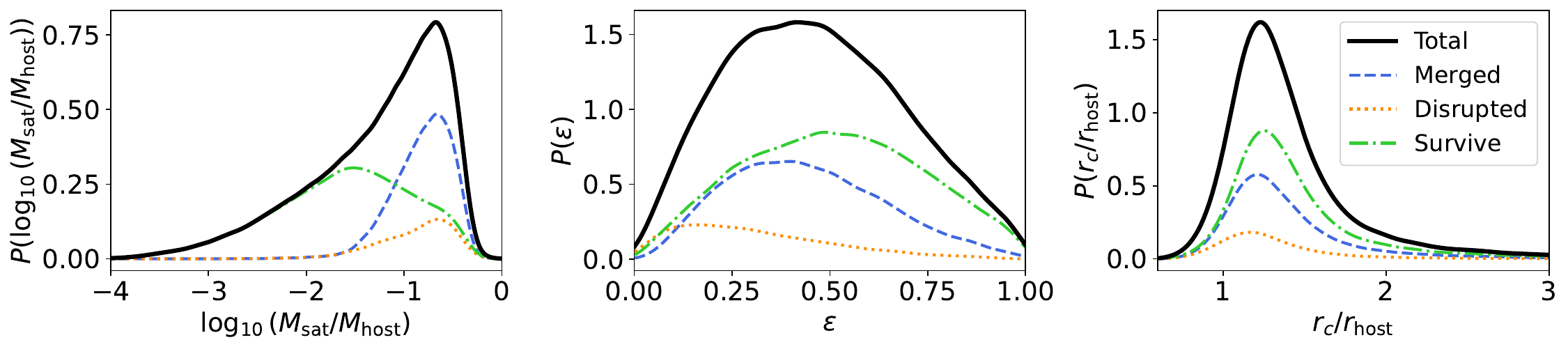}
    \caption{Distributions of \( \log_{10}(M_{\rm{sat}}/M_{\rm{host}}) \), \( \epsilon \), and \( r_c/r_{\rm{host}} \) at the time of first crossing \( r_{\rm{vir}} \) for our entire infall samples, combining \texttt{TNG100-1} and \texttt{TNG300-1}. The distributions are further divided into three sub-samples: merged (\( j_{\rm{merger}}/j_{\rm{infall}} < 0.1 \)), disrupted (\( j_{\rm{merger}}/j_{\rm{infall}} > 0.1 \), regardless of whether the disruption is physical or artificial), and still surviving at \( z = 0 \).}
    \label{fig:infall_dis}
\end{figure*}

Since one of the most significant differences between low- and high-mass galaxies is their morphology—where high-mass galaxies are mostly elliptical while low-mass galaxies are predominantly disk-shaped—we suspect that the difference in merger timescales for low-mass halos is due to the impact of their disk structure. To support this hypothesis, the right panel of Figure~\ref{fig:one2one} presents the logarithmic difference in merger timescales, \(\log_{10}(T_{\rm{dark}}/T_{\rm{baryon}})\), across different \(M_{\rm{host}}\) and \(F_{\rm{disk}}\) bins, where \(F_{\rm{disk}}\) represents the disk-to-total stellar mass fraction of the central galaxy at the merger snapshot \citep{2015ApJ...804L..40G}. Disk particles are identified based on the criterion \(J_{z}/J(E) > 0.7\), where \(J_{z}\) is the specific angular momentum of the particle, and \(J(E)\) is the expected angular momentum for a circular orbit at the same location. We find that \(\log_{10}(T_{\rm{dark}}/T_{\rm{baryon}})\) depends on \(F_{\rm{disk}}\), with larger values observed for higher \(F_{\rm{disk}}\), partially supporting our hypothesis. For high-mass halos, we believe that the resistance of baryons to tidal disruption may be the dominant effect, as many subhalos are disrupted much more quickly in DMO runs. This is evident in the left panel of Figure~\ref{fig:one2one}, where numerous events exhibit \(\log_{10}T_{\rm{dark}} < -0.3\) in the DMO runs but \(\log_{10}T_{\rm{baryon}} > 0\) in the baryonic simulations.

We conduct a more detailed analysis by comparing the infall history of matched merger events in DMO and baryonic runs. Figure~\ref{fig:history} presents two matched events, one in a low-mass host halo hosting a central disk galaxy and the other in a high-mass host halo with a central elliptical galaxy. These two events have similar initial conditions, with \(\epsilon \sim 0.13\) and \(M_{\rm{sat}}/M_{\rm{cen}} \sim 0.1\). The selected events have relatively radial orbits, which exhibit more pronounced differences between the two runs. The mass ratios are chosen to ensure sufficiently long merger histories for clearer illustration, while avoiding rare cases with very small mass ratios.
 Notably, the differences between the DMO and baryonic runs exhibit opposite trends in high-mass and low-mass halos. In the low-mass halo, the subhalo in the baryonic run loses a significant amount of mass after its first pericenter passage and merges with the central galaxy immediately after the second passage. In contrast, in the DMO run, the subhalo survives through four pericenter passages. In the high-mass halo, the subhalo in the DMO run experiences a more rapid loss of mass and sinks to the center faster than in the baryonic run. 

While these analysis supports our hypothesis, further controlled experiments are necessary for a more comprehensive understanding, given the highly chaotic nature of the processes and the numerous influencing factors. Additionally, a thorough investigation is needed to determine whether these effects are physical or merely numerical artifacts.

\begin{figure*}
    \centering
    \includegraphics[width=1\textwidth]{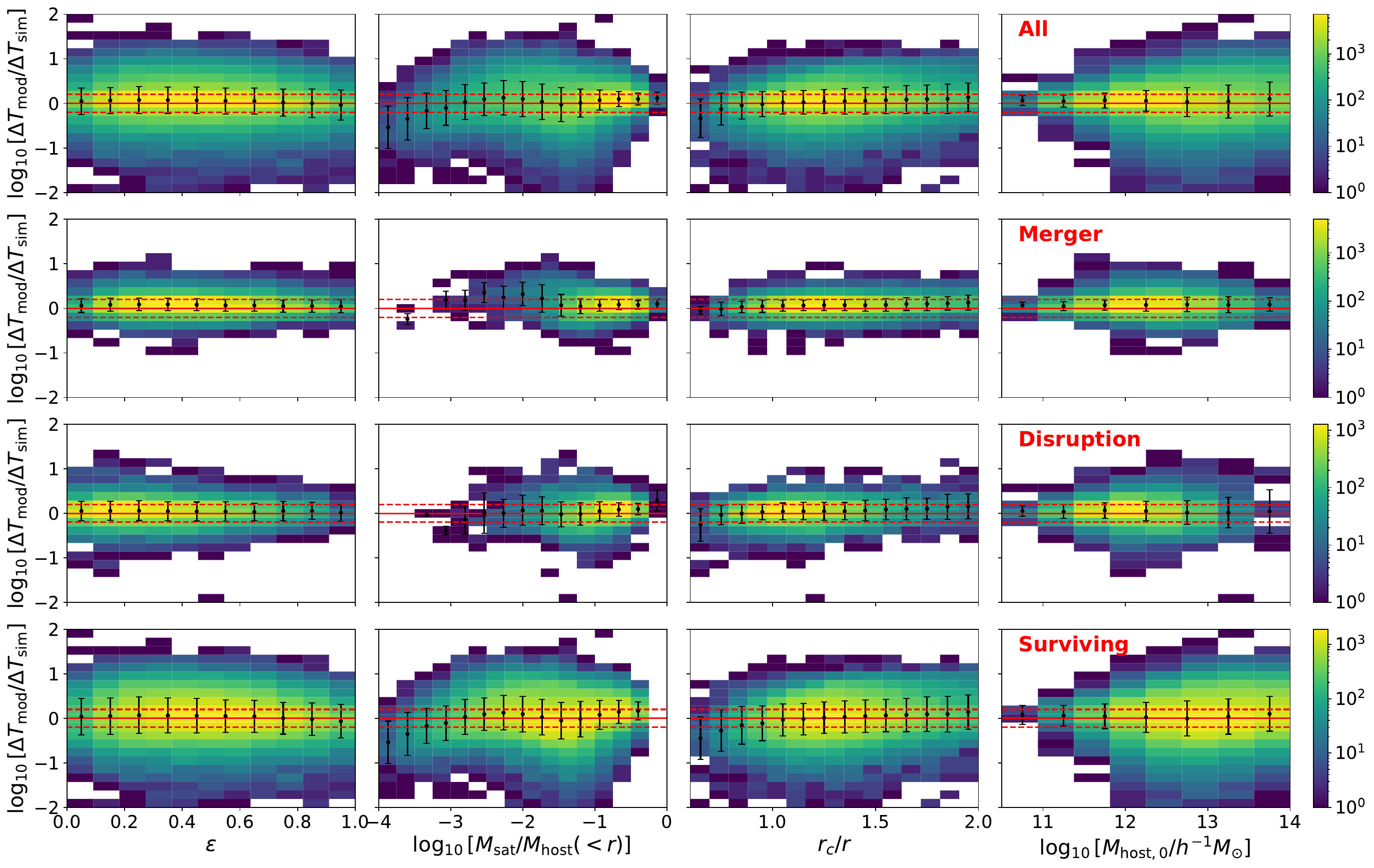}
    \caption{Comparison of \( \Delta T_{\rm merger} \) between simulation results and model predictions using the best-fit model derived from the merger sample started at arbitrary radii (third row of Table \ref{tab:best-fit}). The logarithmic difference, \( \log_{10}[\Delta T_{\rm mod}/\Delta T_{\rm sim}] \), is shown as a function of orbit circularity (\( \epsilon \)), mass ratio (\( M_{\rm sat}/M_{\rm host}(<r) \)), orbit energy (\( r_c/r \)), and host halo mass at the time of crossing \( r_{\rm vir} \) (\( M_{{\rm host},0} \)). Results for the whole infall sample, merger sample, disruption sample, and surviving sample are presented in different rows. The red dashed lines indicate scatter values of -0.2 and 0.2. The figures are color-coded by the number of infall events.}
    \label{fig:extend_test_infall}
\end{figure*}

\begin{figure*}
    \centering
    \includegraphics[width=1\textwidth]{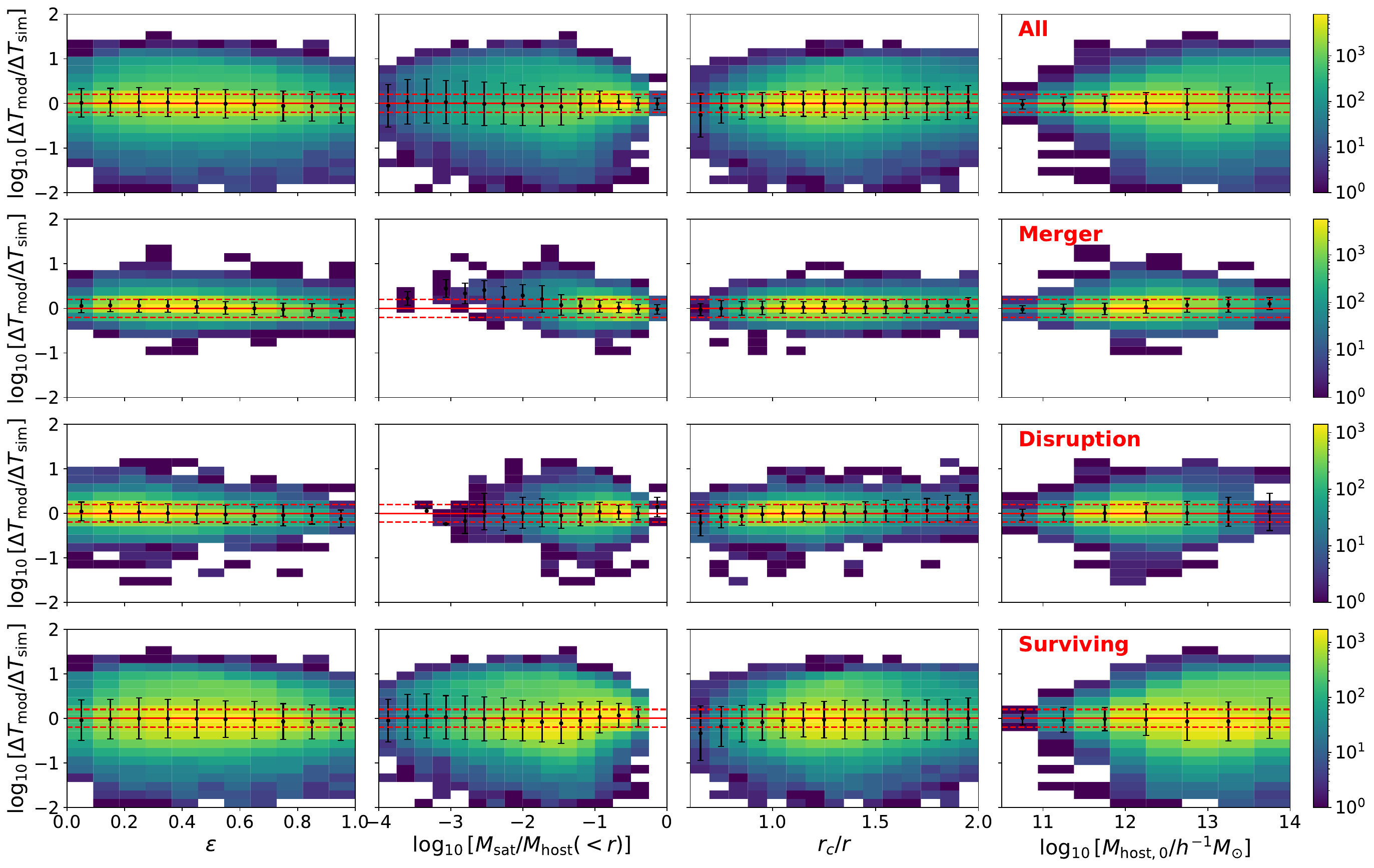}
    \caption{Similar to Figure~\ref{fig:extend_test_infall}, but here the simulation results are compared to the best-fit model derived from fitting the whole infall sample (fourth row of Table \ref{tab:best-fit}). The figures are color-coded by the number of infall events.}
    \label{fig:extend_fit_infall}
\end{figure*}


\section{Models for Arbitrary Starting Radii and selection effects }\label{sec:extend}
In this section, we extend the merger timescale model to include mergers starting at arbitrary starting radii. Using this revised model, we examine selection effects on merger timescale results by comparing the model derived from the merger sample with that from the entire infall sample.

\subsection{Extension to arbitrary starting radii}
Since dark matter halos exhibit universal density profiles \citep{1997ApJ...490..493N}, a natural question arises: can our model be extended to merger events initiated at any radius within the halo, as the inner regions of halos can be considered as new halos with different overdensities (accounted by $T_{\rm dy}$) and concentrations? Some studies have claimed validation for their models \citep{2017MNRAS.472.1392S,2021MNRAS.501.2810P}. Therefore, we extend our model in Equation~\ref{eq:time_scale} to estimate merger timescales for any initial starting radius \( r \) as follows:
    \begin{equation}
    \frac{T_{\rm{merger}}}{T_{\rm{dyn}}(r)} = A e^{\alpha\epsilon} \frac{g[M_{\rm{host}}(<r)/M_{\rm{sat}}]}{\ln[1+M_{\rm{host}}(<r)/M_{\rm{sat}}]} \left[\frac{r_c(E)}{r}\right]^{\gamma}\,, \label{eq:time_scale_ex}
\end{equation}  
where \( M_{\rm{host}}(<r) \) is the mass of the halo enclosed within radius \( r \), \( T_{\rm{dyn}}(r) = \sqrt{r^3 / [G M_{\rm{host}}(<r)]} \) is the dynamical time, \( M_{\rm{sat}} \) represents the bound mass of the subhalo, and the orbital parameters \( \epsilon \) and \( r_c \) are still computed using the SIS potential for convenience, constrained by \( (r, M_{\rm{host}}(<r)) \). We first maintain the mass ratio dependence as \( g(x) = x^{\beta} \). The mass dependence of \( \alpha \), \( \beta \), and \( \gamma \) is preserved as a function of the virial mass at the first crossing of $r_{\rm vir}$, following Equation~\ref{eq:mass_dependence}, with the best-fit values given in the first row of Table \ref{tab:best-fit}.

To test the model, we use the merger sample from \texttt{TNG100-1} and \texttt{TNG300-1}. For each event, we treat each intermediate snapshot during its infall history as the starting point of a new event and compute the corresponding initial conditions and merger timescales. This process requires particle data from each snapshot to determine \( M_{\rm{host}}(<r) \). In Figure~\ref{fig:extend_test}, we present the logarithmic difference, \( \log_{10}(T_{\rm{mod}}/T_{\rm{sim}}) \), between simulation results and model predictions from Equation~\ref{eq:time_scale_ex}. This difference is shown as a function of \(\epsilon\), mass ratio \(M_{\rm sat}/M_{\rm host}(<r)\), orbit energy \(r_c/r\), host halo mass \(M_{{\rm host},0}\) at the time of crossing \( r_{\rm host,0}=r_{\rm vir} \), and starting radius \(r\). We find that after entering the halo (\( \log_{10}(r/r_{\rm host,0}) < -0.3 \)), our extended model underestimates the merger timescales, suggesting a transition before and after infall. Moreover, our model significantly overestimates the merger timescales for mass ratios \( <10^{-2} \), where fewer cases were present in the previous sample used for parameter constraints. This may be reasonable, as  subhalos in low mass ratio events experience weaker dynamical friction but are more susceptible to disruption. As a result, this effect may cause events with low mass ratio to deviate from the previous mass dependence.

To better describe the merger timescales for events starting at any radius within the halo, we introduce a correction based on $r$ and adjust the mass ratio dependence at the low-mass end. We find that the dependence of \( \log_{10}(T_{\rm{mod}}/T_{\rm{sim}}) \) on \( r \) originates from \( M_{\rm host}(<r) \). Figure~\ref{fig:mass_correct} shows \( M_{\rm host}(<r)/M_{\rm host,0} \) as a function of \( r/r_{\rm host,0} \) for our merger sample. The fraction \( M_{\rm host}(<r)/M_{\rm host,0} \) increases with \( r/r_{\rm host,0} \), following a power-law trend up to \( r/r_{\rm host,0} \sim 10^{-0.5} \), beyond which the growth gradually slows down. This is due to the halo's non-spherical mass distribution and boundary. As \( r \) approaches \( r_{\rm vir} \), the enclosed sphere extends beyond the irregular halo boundary, capturing little additional mass and slowing the total mass increase. If \( M_{\rm host}(<r) \) determines the merger timescale, this is not an issue. However, if the relevant factor is not the enclosed spherical mass but an alternative effective mass, such as the elliptical mass, this trend could introduce a bias across different \( r \), as demonstrated in Figure~\ref{fig:extend_test}. This bias can be corrected by defining an effective host halo mass. The relationship between \( M_{\rm host}(<r)/M_{\rm host,0} \) and \( r/r_{\rm host,0} \) is well described by  
\begin{equation}  
\frac{M_{\rm host}(<r)}{M_{\rm host,0}} = \frac{1 + r_{\rm t}^{-\eta}}{(r/r_{\rm host,0})^{-\eta} + r_{\rm t}^{-\eta}},  
\label{eq:mass_correct}  
\end{equation}  
with best-fit parameters \( r_{\rm t} = 0.833 \) and \( \eta = 1.257 \). Based on this relation, we define a correction factor  
\begin{equation}  
f_c(r/r_{\rm host,0}) = \frac{(r/r_{\rm host,0})^{\eta} + r_{\rm t}^{\eta}}{1 + r_{\rm t}^{\eta}},  
\label{eq:factor}  
\end{equation}  
which is normalized such that \( f_c(1) = 1 \), ensuring consistency with our previous results. The effective host halo mass is then defined as \( f_c M_{\rm host}(<r) \), which is also presented in Figure~\ref{fig:mass_correct}. To better capture the mass ratio dependence, motivated by the power-law upturn bias seen in Figure~\ref{fig:extend_test}, we replace $[M_{\rm{host}}(<r)/M_{\rm{sat}}]^{\beta}$ with a double power-law form:  
\begin{equation}  
g(M_{\rm ratio})=\frac{M_{\beta}^{\beta_1} + M_{\beta}^{\beta_2}}{(M_{\rm{ratio}}/M_{\beta})^{-\beta_1} + (M_{\rm{ratio}}/M_{\beta})^{-\beta_2}} \,,  \label{eq:mass_double}
\end{equation}  
where \( M_{\rm{ratio}} = M_{\rm{host}}(<r) / M_{\rm{sat}} \). We test replacing \( M_{\rm host}(<r) \) with \( f_c M_{\rm host}(<r) \) in \( M_{\rm ratio} \), \( \Lambda \), and \( T_{\rm dy} \) and find that only \( T_{\rm dy} \) requires modification. Adjusting the other components have negligible or adverse effects on the results. Therefore, we define a new $T_{\rm dy}$ as
\begin{equation}
    T_{\rm dy}=\sqrt{\frac{r^3}{Gf_cM_{\rm host}(<r)}}\,.\label{eq:Tdy}
\end{equation}

By combining Equations \ref{eq:mass_dependence}, \ref{eq:time_scale_ex}, \ref{eq:factor},  \ref{eq:mass_double}, and \ref{eq:Tdy}, we develop a model to describe merger timescales for mergers starting at arbitrary starting radii. This model includes 11 parameters: \( \{A_{\rm{h}}, M_{\rm{min}}, \sigma_{\log M}, A_{\rm{min}}, \alpha_{\rm{h}}, \alpha_{\rm{min}}, \gamma_{\rm{h}}, \gamma_{\rm{min}}, \beta_1, \beta_2, M_{\beta} \} \). When including \( r_{\rm t} \) and \( \eta \), the total number of parameters increases to 13. We fit this new model to the merger sample starting at arbitrary starting radii. The best-fit parameters are listed in the third row of Table \ref{tab:best-fit}, and the residuals as a function of various parameters are shown in Figure~\ref{fig:extend_fit}. Our new model effectively describes the merger timescales for the arbitrary starting radius sample, achieving \(\sigma_{\log} = 0.195\), with the scatter showing little to no dependence on any parameters. In Figure~\ref{fig:extend_fit}, we also show the scatter for the best-fit model without the correction for \( T_{\rm dy} \) in the \( r/r_{\rm host,0} \) panel. Including this correction really helps reduce the bias dependence on \( r/r_{\rm host,0} \). In the new model, we find a weaker dependence of \( \alpha \) and \( \gamma \) on \( M_{\rm host,0} \) compared to the previous model, with \( \alpha_h = 0.406 \) (previously 0.537) and \( \gamma_h = -0.026 \) (previously 0.193). Notably, \( \gamma \) now shows almost no dependence. This suggests that \( M_{\rm host,0} \) may not be a reliable indicator of system variation for mergers that do not start at \( r_{\rm vir} \), and more fundamental parameters might be needed. However, since our model already performs well and \( M_{\rm host,0} \) remains the most accessible quantity, we do not explore alternative options.

\subsection{The entire infall sample and selection effects}\label{sec:5.2}
One remaining concern regarding our merger timescale results is selection effects, a common issue in studies of merger timescales using cosmological simulations. Since we only analyze satellites that have already merged with centrals, our sample is likely biased toward events with shorter merger timescales. If this sample is taken as representative of all infall events, our model may underestimate the true average merger timescales. With the model for arbitrary starting radii, we can now use the entire infall sample—whether the satellites have merged, been disrupted, or are still surviving—to study selection effects. This is possible because our model allows us to predict merger timescales for any two points in an event's timeline and compare the predicted time intervals with those from simulations. As a result, satellites do not need to have already merged with the central to be included in the analysis.

To construct the infall sample, we select all satellites that have crossed \( r_{\rm vir} \) after \( z = 4 \), using only the mass selection criterion as before: \( M_{\rm{vir}}(z=0) > 10^{10.7} h^{-1} M_{\odot} \) and \( M_{*}^{\rm sat}(z_{\rm infall}) > 10^{8.3} h^{-1} M_{\odot} \) for \texttt{TNG100-1}, and \( M_{\rm{vir}}(z=0) > 10^{11.7} h^{-1} M_{\odot} \) and \( M_{*}^{\rm sat}(z_{\rm infall}) > 10^{9.3} h^{-1} M_{\odot} \) for \texttt{TNG300-1}. In Figure~\ref{fig:infall_dis}, we present the distributions of the initial properties—\( \log_{10}(M_{\rm sat}/M_{\rm host}) \), \( \epsilon \), and \( r_c/r_{\rm host} \)—at the time of first crossing \( r_{\rm vir} \) for the infall sample. We also show the contributions from three sub-samples: satellites that have merged with the centrals (\( \log_{10}(j_{\rm merger}/j_{\rm infall}) < -1 \)), those that have been disrupted (\( \log_{10}(j_{\rm merger}/j_{\rm infall}) > -1 \), regardless of whether the disruption is physical or artificial), and those that are still surviving at \( z = 0 \). If we use only the merger sample instead of the full infall sample as before, we tend to select satellites with higher mass ratios, more radial orbits , and possibly shorter merger timescale events for the same initial conditions, which may bias our model toward underestimating merger timescales.

Therefore, we refine our model by incorporating the entire infall sample. For the merger sample, we maintain the previous approach, requiring only the prediction of the merger timescale at the infall time, \( T_{\rm merger}(t_{\rm infall}) \), where \( t_{\rm infall} \) is defined as the time of first crossing \( r_{\rm vir} \). For the disruption and surviving samples, we predict both \( T_{\rm merger}(t_{\rm infall}) \) and \( T_{\rm merger}(t_{50}) \), where \( t_{50} \) is the last time the subhalo contained more than 50 particles. If the subhalo still has more than 50 particles at \( z=0 \), then \( t_{50} \) corresponds to the present time. The requirement of at least 50 remaining particles is imposed to reduce numerical artifacts. With these predictions, we compute \( \Delta T_{\rm merger} = T_{\rm merger}(t_{\rm infall}) - T_{\rm merger}(t_{50}) \) from our model and compare it with simulation results. For the merger sample, \( \Delta T_{\rm merger} \) simply equals \( T_{\rm merger}(t_{\rm infall}) \). We have also tested $t_{100}$ and found similar results, confirming the robustness of this choice.

We first assess whether our previously derived arbitrary starting radii model, based on the merger sample, accurately predicts \( \Delta T_{\rm merger} \) for the entire infall sample. Figure~\ref{fig:extend_test_infall} compares model predictions of \( \Delta T_{\rm merger} \) with simulation results. Overall, the model performs well across all three sub-samples, except when the mass ratio is very small (\( M_{\rm sat}/M_{\rm host} < 10^{-3} \)). For the merger sample, \( T_{\rm merger} \) from simulations is generally slightly shorter than the model predictions, with discrepancies reaching ~0.2 dex for \( M_{\rm sat}/M_{\rm host} < 10^{-2} \). This aligns with expectations, as the merger sample tends to favor shorter merger timescales. Despite being calibrated on the merger sample, the model’s ability to incorporate information from arbitrary radii appears to mitigate selection effects, likely due to the collisional nature of the merger process.

We refine our model by fitting \( \Delta T_{\rm merger} \) from the entire infall sample, using the same framework as the arbitrary starting radii model. The best-fit parameters are listed in the fourth row of Table \ref{tab:best-fit}, and Figure~\ref{fig:extend_fit_infall} compares model predictions with simulation results. The new model shows improved accuracy, particularly for low mass ratios (\( M_{\rm sat}/M_{\rm host} < 10^{-3} \)) and high \( r_c/r \), with an average logarithmic scatter of \( \sigma_{\log} = 0.317 \) for the entire infall sample. For \( M_{\rm sat}/M_{\rm host} < 10^{-2} \), \( T_{\rm merger} \) from the merger sample remains 0.2 dex lower than the model prediction, highlighting the presence of selection effects. This also indicates that incorporating the entire infall sample into the fitting framework effectively reduces selection effects. However, this comes at the cost of increased scatter due to the need for predictions at two different times.

\section{Conclusion and Discussion}\label{sec:con}
In this paper, we investigate galaxy merger timescales in the \texttt{IllustrisTNG} simulations and develop models that predict merger timescales based on the initial conditions—\( M_{\rm{sat}}/M_{\rm{host}} \), \( \epsilon \), and \( r_c(E)/r_{\rm{host}} \)—when the satellite first crosses \( r_{\rm{vir}} \). Our model is further extended to describe merger events initiated at arbitrary radii within the host halo. Additionally, we examine the impact of halo formation history (characterized by halo mass), baryonic effects, and sample selection on merger timescales. We summarize our results as follows.
\begin{itemize}
    \item Using merger samples from \texttt{TNG100-1} and \texttt{TNG300-1}, we find that the dependence of merger timescales on initial conditions is not universal, with massive halos exhibiting a weaker dependence on orbital parameters. This may be due to more violent processes, such as major mergers and subhalo interactions, which make the infall process more collisional and weaken its connection to initial conditions. To account for this, we introduce an \( M_{\rm{host}} \)-dependent model that accurately predicts merger timescales across host halos of different masses.
    \item Compared to baryonic runs, merger timescales in DMO runs are similar for circular orbits but show larger differences for radial orbits. In low-mass halos (\( M_{\rm{host}} < 10^{12.5} h^{-1} M_{\odot} \)), mergers in DMO runs have longer timescales than in baryonic runs, whereas in high-mass halos (\( M_{\rm{host}} > 10^{12.5} h^{-1} M_{\odot} \)), the trend reverses. This may be because, in low-mass halos, the central galaxy's high \( F_{\rm{disk}} \) leads to a disk structure that more effectively disrupts satellites. In high-mass halos, however, the resistance of baryons within satellites to tidal disruption may become the dominant effect.
    \item With a slight modification to the mass ratio dependence, our model can be extended to merger events originating from any radius within the halo. Using this new model, we investigate selection effects on merger timescale predictions by incorporating the entire infall sample into the fitting, rather than just the merger sample. We find that relying solely on the merger sample leads to an underestimation of average merger timescales, particularly for low-mass-ratio events with \( M_{\rm sat}/M_{\rm host} < 10^{-2} \). In contrast, our model, derived from the full infall sample, provides a more unbiased prediction.
    \item Table \ref{tab:best-fit} presents four best-fit models for merger timescales, each derived from different samples. We recommend using the fourth model, which is based on the entire infall sample and applicable to any starting radius. The other models may still be useful for specific cases.
\end{itemize}

Our results indicate that merger timescales may be influenced by the formation history of halos and baryonic effects, potentially explaining the discrepancies among previous studies that examined different types of systems. Additionally, selection effects should be carefully considered in related studies to ensure accurate predictions. 

Our models for galaxy merger timescales, presented in Table \ref{tab:best-fit}, serve as valuable tools for semi-analytical and empirical models of galaxy formation. The evolution of subhalos (satellites) is a crucial component of hierarchical structure formation. Unlike halo formation and evolution—where physically motivated semi-analytical models such as the Press-Schechter framework \citep{1974ApJ...187..425P,1991ApJ...379..440B,1993MNRAS.262..627L} provide theoretical guidance—subhalo evolution after halo mergers is highly non-linear and chaotic, making it challenging to model without numerical experiments \citep{2001ApJ...559..716T,2004ApJ...608..663K,2016MNRAS.457.1208H,2022MNRAS.516..106D,2023MNRAS.521.4432S}. The merger timescale is a key parameter in determining subhalo abundance, making our models an essential input for semi-analytical and empirical approaches to galaxy formation. Moreover, even in high-resolution numerical simulations, halos are more resilient to numerical artifacts than subhalos. The abundance and distribution of halos can converge with as few as 10–20 particles \citep{2019SCPMA..6219511J}. However, for subhalos, the high-density environment within host halos leads to rapid tidal disruption, requiring at least 50 particles for a robust simulation of subhalo evolution \citep{2018MNRAS.475.4066V,2025ApJ...981..108H}, which can be computationally inefficient. In practice, the most bound particles of artificially disrupted subhalos are often tracked to approximate their unresolved evolution, but their survival time must be estimated carefully, as these particles may experience significantly weaker dynamical friction than fully resolved subhalos. If merger timescales are not considered when determining subhalo survival, their abundance may be overestimated. Thus, our galaxy merger timescale models play a crucial role in improving the accuracy of galaxy formation modeling.

Our merger timescale model is also useful for inferring galaxy merger rates from observed close pairs \citep{2008MNRAS.391.1489K,2014ApJ...790....7J}, which play a key role in the ex situ growth of galaxies. Estimating galaxy merger rates typically involves identifying galaxy pairs and determining merger timescales as a function of separation distance, where our model with arbitrary starting radii can provide valuable input \citep{2020ApJ...895..115F,2021ApJ...919..139W}. Moreover, the galaxy merger timescales defined in our study effectively represent the delay between halo mergers and the corresponding galaxy mergers. As such, they can be useful for inferring the halo merger rate, which is not directly observable but is closely linked to cosmological parameters \citep{2024MNRAS.527.3459A}.

\section*{Acknowledgments}
We thank the anonymous referee for the detailed and valuable comments that helped improve the manuscript. We thank Jiaxin Han, Shaun Cole, and Carlos Frenk for their valuable discussions. We also appreciate Dylan Nelson and Vicente Rodriguez-Gomez for their assistance in accessing the merger trees of \texttt{IllustrisTNG}. K.X. is supported by the funding from the Center for Particle Cosmology at U Penn. Y.P.J. is supported by NSFC (12133006, 11890691), by National Key R\&D Program of China (2023YFA1607800, 2023YFA1607801), by grant No. CMS-CSST-2021-A03, and by 111 project No. B20019. This work made use of the Gravity Supercomputer at the Department of Astronomy, Shanghai Jiao Tong University.

%






\appendix
\restartappendixnumbering
\section{Validation of fits}\label{sec:appendix}
\begin{figure*}
    \centering
    \includegraphics[width=1\textwidth]{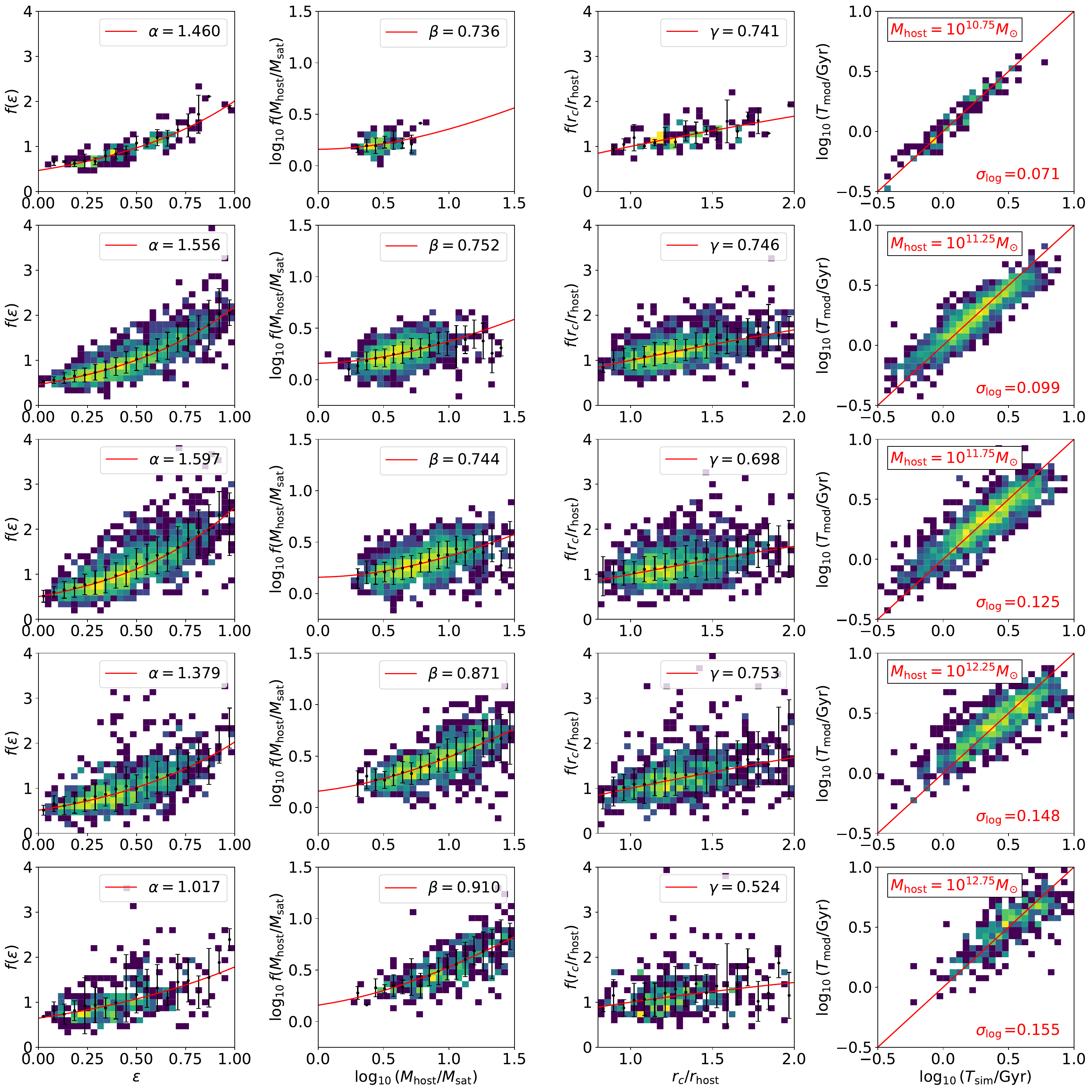}
    \caption{Similar to Figure~\ref{fig:fit_full}, but for the \texttt{TNG100-1} sample divided into different $M_{\rm{host}}$ bins.}
    \label{fig:fit_tng100}
\end{figure*}

\begin{figure*}
    \centering
    \includegraphics[width=1\textwidth]{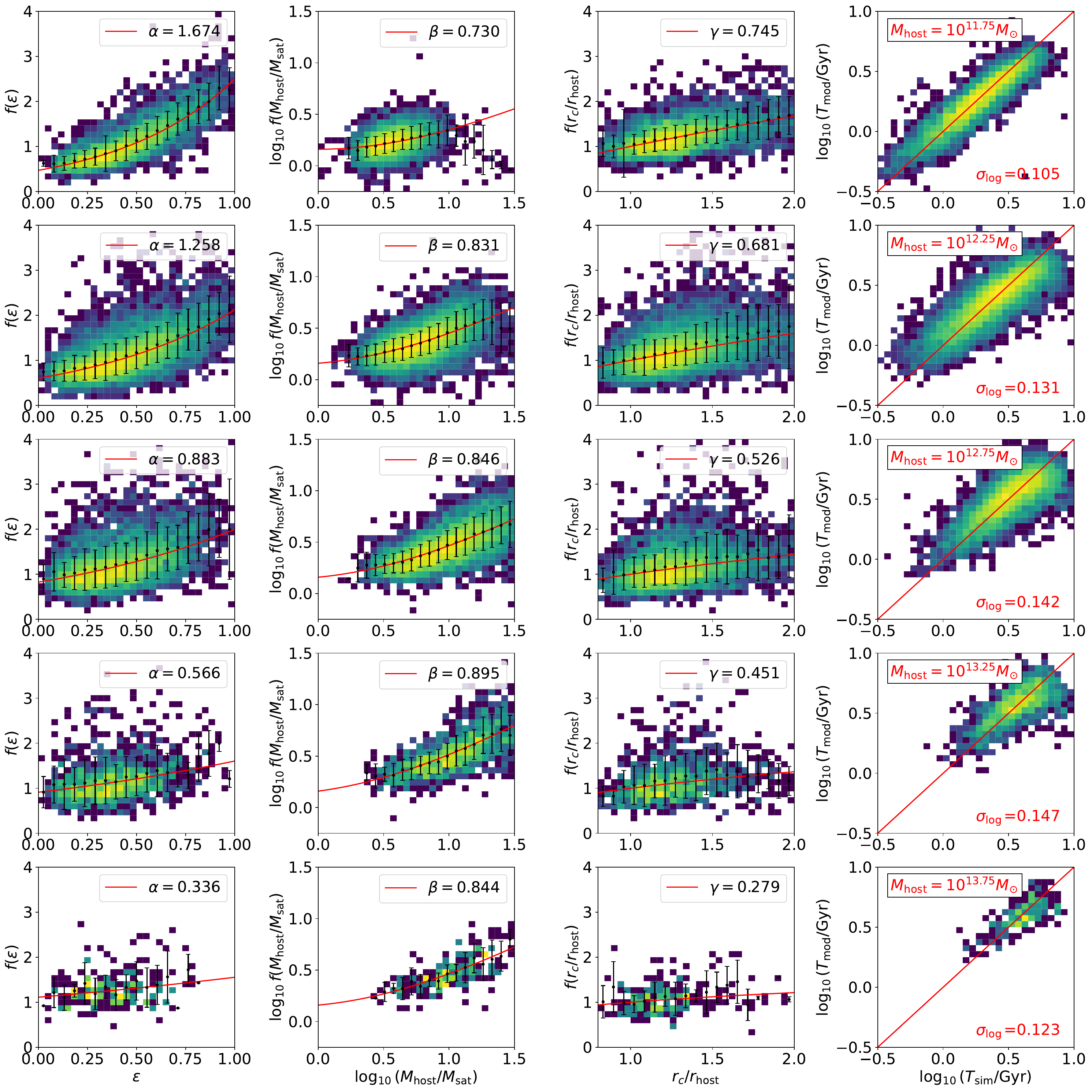}
    \caption{Similar to Figure~\ref{fig:fit_full}, but for the \texttt{TNG300-1} sample divided into different $M_{\rm{host}}$ bins.}
    \label{fig:fit_tng300}
\end{figure*}

\begin{figure*}
    \centering
    \includegraphics[width=1\textwidth]{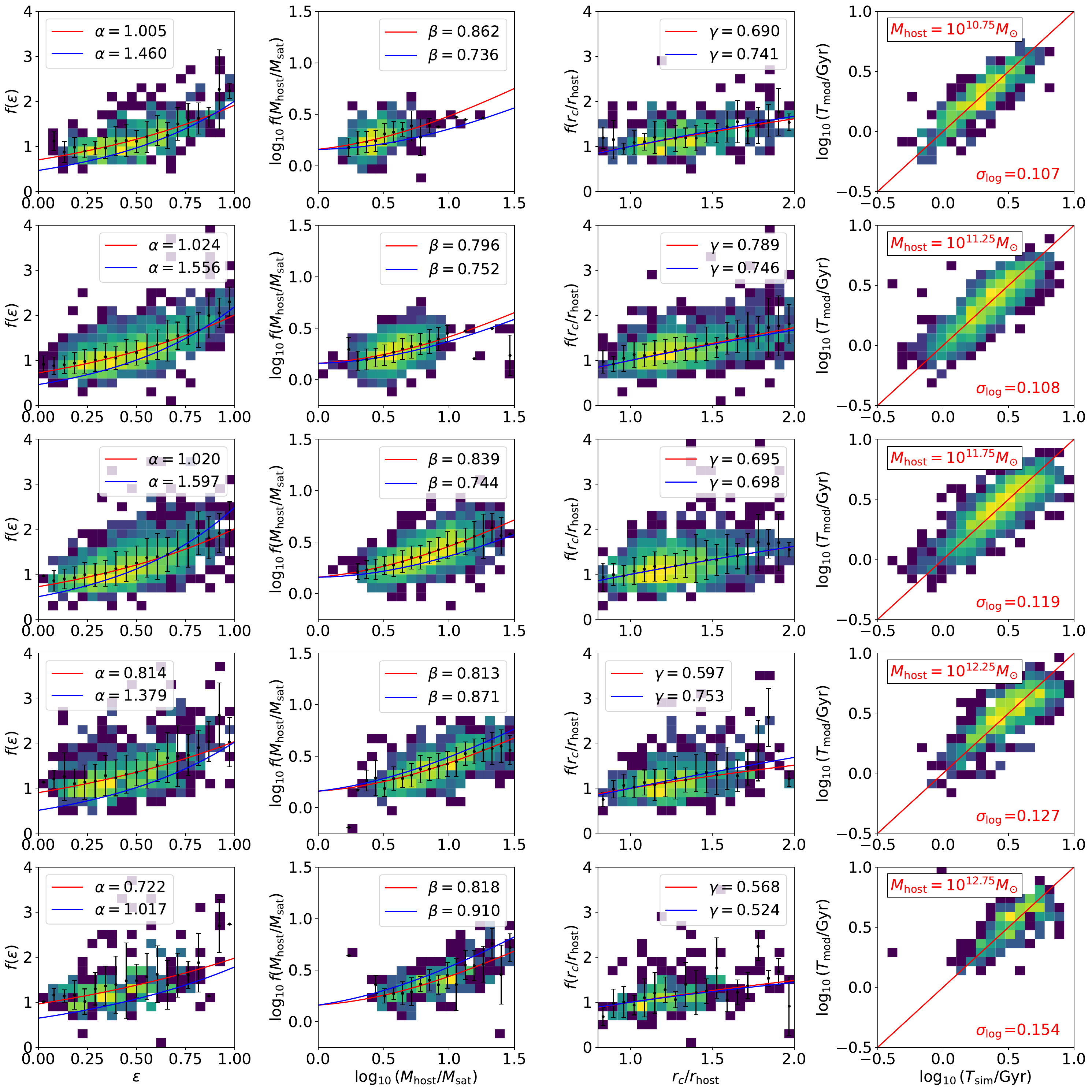}
    \caption{Similar to Figure~\ref{fig:fit_full}, but for the \texttt{TNG100-1-Dark} sample divided into different $M_{\rm{host}}$ bins. The best-fit models from \texttt{TNG100-1} are also shown in blue lines for comparison.}
    \label{fig:fit_tng100_dark}
\end{figure*}

\begin{figure*}
    \centering
    \includegraphics[width=1\textwidth]{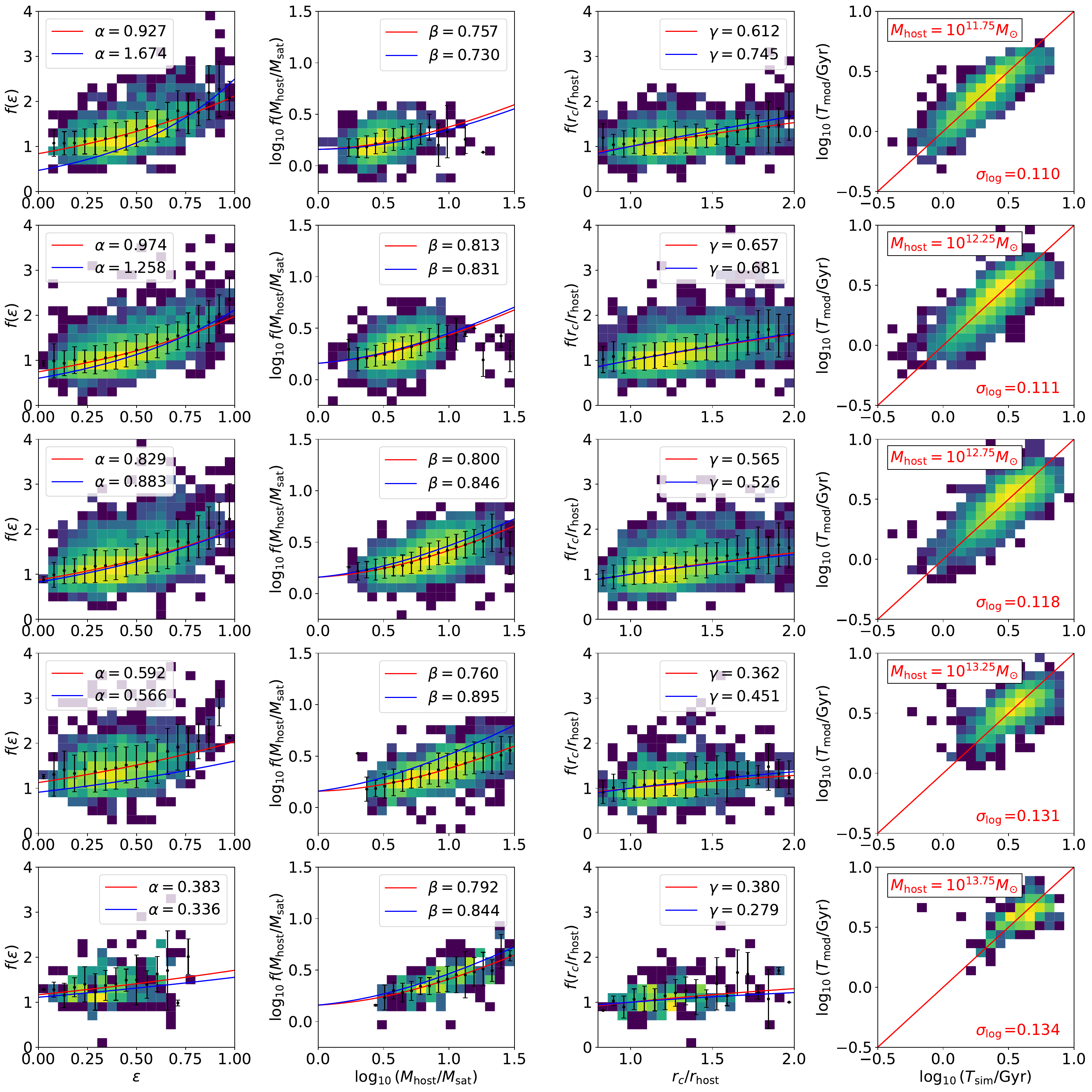}
    \caption{Similar to Figure~\ref{fig:fit_full}, but for the \texttt{TNG300-1-Dark} sample divided into different $M_{\rm{host}}$ bins. The best-fit models from \texttt{TNG300-1} are also shown in blue lines for comparison.}
    \label{fig:fit_tng300_dark}
\end{figure*}

In this section, we validate the merger timescale fits across different \( M_{\rm host} \) bins. Results for \texttt{TNG100-1}, \texttt{TNG300-1}, \texttt{TNG100-1-Dark}, and \texttt{TNG300-1-Dark} are shown in Figures~\ref{fig:fit_tng100}, \ref{fig:fit_tng300}, \ref{fig:fit_tng100_dark}, and \ref{fig:fit_tng300_dark}, respectively.

\bibliography{sample631}{}

\begin{thebibliography}{}
\expandafter\ifx\csname natexlab\endcsname\relax\def\natexlab#1{#1}\fi
\providecommand{\url}[1]{\href{#1}{#1}}
\providecommand{\dodoi}[1]{doi:~\href{http://doi.org/#1}{\nolinkurl{#1}}}
\providecommand{\doeprint}[1]{\href{http://ascl.net/#1}{\nolinkurl{http://ascl.net/#1}}}
\providecommand{\doarXiv}[1]{\href{https://arxiv.org/abs/#1}{\nolinkurl{https://arxiv.org/abs/#1}}}

\bibitem[{{Amoura} {et~al.}(2024){Amoura}, {Drakos}, {Berrouet}, \&
  {Taylor}}]{2024MNRAS.527.3459A}
{Amoura}, Y., {Drakos}, N.~E., {Berrouet}, A., \& {Taylor}, J.~E. 2024, \mnras,
  527, 3459, \dodoi{10.1093/mnras/stad3416}

\bibitem[{{Binney} \& {Tremaine}(1987)}]{1987gady.book.....B}
{Binney}, J., \& {Tremaine}, S. 1987, {Galactic dynamics} (Princeton University
  Press)

\bibitem[{{Bond} {et~al.}(1991){Bond}, {Cole}, {Efstathiou}, \&
  {Kaiser}}]{1991ApJ...379..440B}
{Bond}, J.~R., {Cole}, S., {Efstathiou}, G., \& {Kaiser}, N. 1991, \apj, 379,
  440, \dodoi{10.1086/170520}

\bibitem[{{Boylan-Kolchin} {et~al.}(2008){Boylan-Kolchin}, {Ma}, \&
  {Quataert}}]{2008MNRAS.383...93B}
{Boylan-Kolchin}, M., {Ma}, C.-P., \& {Quataert}, E. 2008, \mnras, 383, 93,
  \dodoi{10.1111/j.1365-2966.2007.12530.x}

\bibitem[{{Bryan} \& {Norman}(1998)}]{1998ApJ...495...80B}
{Bryan}, G.~L., \& {Norman}, M.~L. 1998, \apj, 495, 80, \dodoi{10.1086/305262}

\bibitem[{{Chandrasekhar}(1943)}]{1943ApJ....97..255C}
{Chandrasekhar}, S. 1943, \apj, 97, 255, \dodoi{10.1086/144517}

\bibitem[{{Cole} {et~al.}(2000){Cole}, {Lacey}, {Baugh}, \&
  {Frenk}}]{2000MNRAS.319..168C}
{Cole}, S., {Lacey}, C.~G., {Baugh}, C.~M., \& {Frenk}, C.~S. 2000, \mnras,
  319, 168, \dodoi{10.1046/j.1365-8711.2000.03879.x}

\bibitem[{{Colpi} {et~al.}(1999){Colpi}, {Mayer}, \&
  {Governato}}]{1999ApJ...525..720C}
{Colpi}, M., {Mayer}, L., \& {Governato}, F. 1999, \apj, 525, 720,
  \dodoi{10.1086/307952}

\bibitem[{{De Lucia} \& {Blaizot}(2007)}]{2007MNRAS.375....2D}
{De Lucia}, G., \& {Blaizot}, J. 2007, \mnras, 375, 2,
  \dodoi{10.1111/j.1365-2966.2006.11287.x}

\bibitem[{{Delfino} {et~al.}(2022){Delfino}, {Sc{\'o}ccola}, {Cora},
  {Vega-Mart{\'\i}nez}, \& {Gargiulo}}]{2022MNRAS.510.2900D}
{Delfino}, F.~M., {Sc{\'o}ccola}, C.~G., {Cora}, S.~A., {Vega-Mart{\'\i}nez},
  C.~A., \& {Gargiulo}, I.~D. 2022, \mnras, 510, 2900,
  \dodoi{10.1093/mnras/stab3494}

\bibitem[{{Drakos} {et~al.}(2022){Drakos}, {Taylor}, \&
  {Benson}}]{2022MNRAS.516..106D}
{Drakos}, N.~E., {Taylor}, J.~E., \& {Benson}, A.~J. 2022, \mnras, 516, 106,
  \dodoi{10.1093/mnras/stac2202}

\bibitem[{{Drakos} {et~al.}(2019{\natexlab{a}}){Drakos}, {Taylor}, {Berrouet},
  {Robotham}, \& {Power}}]{2019MNRAS.487..993D}
{Drakos}, N.~E., {Taylor}, J.~E., {Berrouet}, A., {Robotham}, A. S.~G., \&
  {Power}, C. 2019{\natexlab{a}}, \mnras, 487, 993,
  \dodoi{10.1093/mnras/stz1306}

\bibitem[{{Drakos} {et~al.}(2019{\natexlab{b}}){Drakos}, {Taylor}, {Berrouet},
  {Robotham}, \& {Power}}]{2019MNRAS.487.1008D}
---. 2019{\natexlab{b}}, \mnras, 487, 1008, \dodoi{10.1093/mnras/stz1307}

\bibitem[{{Du} {et~al.}(2024){Du}, {Benson}, {Zeng}, {Treu}, {Peter}, {Mace},
  {Jiang}, {Yang}, {Gannon}, {Gilman}, {Nierenberg}, \&
  {Nadler}}]{2024PhRvD.110b3019D}
{Du}, X., {Benson}, A., {Zeng}, Z.~C., {et~al.} 2024, \prd, 110, 023019,
  \dodoi{10.1103/PhysRevD.110.023019}

\bibitem[{{Ferreira} {et~al.}(2020){Ferreira}, {Conselice}, {Duncan}, {Cheng},
  {Griffiths}, \& {Whitney}}]{2020ApJ...895..115F}
{Ferreira}, L., {Conselice}, C.~J., {Duncan}, K., {et~al.} 2020, \apj, 895,
  115, \dodoi{10.3847/1538-4357/ab8f9b}

\bibitem[{{Garrison-Kimmel} {et~al.}(2017){Garrison-Kimmel}, {Wetzel},
  {Bullock}, {Hopkins}, {Boylan-Kolchin}, {Faucher-Gigu{\`e}re}, {Kere{\v{s}}},
  {Quataert}, {Sanderson}, {Graus}, \& {Kelley}}]{2017MNRAS.471.1709G}
{Garrison-Kimmel}, S., {Wetzel}, A., {Bullock}, J.~S., {et~al.} 2017, \mnras,
  471, 1709, \dodoi{10.1093/mnras/stx1710}

\bibitem[{{Genel} {et~al.}(2015){Genel}, {Fall}, {Hernquist}, {Vogelsberger},
  {Snyder}, {Rodriguez-Gomez}, {Sijacki}, \& {Springel}}]{2015ApJ...804L..40G}
{Genel}, S., {Fall}, S.~M., {Hernquist}, L., {et~al.} 2015, \apjl, 804, L40,
  \dodoi{10.1088/2041-8205/804/2/L40}

\bibitem[{{Guo} {et~al.}(2011){Guo}, {White}, {Boylan-Kolchin}, {De Lucia},
  {Kauffmann}, {Lemson}, {Li}, {Springel}, \& {Weinmann}}]{2011MNRAS.413..101G}
{Guo}, Q., {White}, S., {Boylan-Kolchin}, M., {et~al.} 2011, \mnras, 413, 101,
  \dodoi{10.1111/j.1365-2966.2010.18114.x}

\bibitem[{{Han} {et~al.}(2016){Han}, {Cole}, {Frenk}, \&
  {Jing}}]{2016MNRAS.457.1208H}
{Han}, J., {Cole}, S., {Frenk}, C.~S., \& {Jing}, Y. 2016, \mnras, 457, 1208,
  \dodoi{10.1093/mnras/stv2900}

\bibitem[{{He} {et~al.}(2025){He}, {Han}, \& {Li}}]{2025ApJ...981..108H}
{He}, F., {Han}, J., \& {Li}, Z. 2025, \apj, 981, 108,
  \dodoi{10.3847/1538-4357/adb383}

\bibitem[{{Helmi} {et~al.}(2003){Helmi}, {Navarro}, {Meza}, {Steinmetz}, \&
  {Eke}}]{2003ApJ...592L..25H}
{Helmi}, A., {Navarro}, J.~F., {Meza}, A., {Steinmetz}, M., \& {Eke}, V.~R.
  2003, \apjl, 592, L25, \dodoi{10.1086/377364}

\bibitem[{{Henriques} {et~al.}(2015){Henriques}, {White}, {Thomas}, {Angulo},
  {Guo}, {Lemson}, {Springel}, \& {Overzier}}]{2015MNRAS.451.2663H}
{Henriques}, B. M.~B., {White}, S. D.~M., {Thomas}, P.~A., {et~al.} 2015,
  \mnras, 451, 2663, \dodoi{10.1093/mnras/stv705}

\bibitem[{{Hopkins} {et~al.}(2006){Hopkins}, {Hernquist}, {Cox}, {Di Matteo},
  {Robertson}, \& {Springel}}]{2006ApJS..163....1H}
{Hopkins}, P.~F., {Hernquist}, L., {Cox}, T.~J., {et~al.} 2006, \apjs, 163, 1,
  \dodoi{10.1086/499298}

\bibitem[{{Jiang} {et~al.}(2008){Jiang}, {Jing}, {Faltenbacher}, {Lin}, \&
  {Li}}]{2008ApJ...675.1095J}
{Jiang}, C.~Y., {Jing}, Y.~P., {Faltenbacher}, A., {Lin}, W.~P., \& {Li}, C.
  2008, \apj, 675, 1095, \dodoi{10.1086/526412}

\bibitem[{{Jiang} {et~al.}(2014){Jiang}, {Jing}, \&
  {Han}}]{2014ApJ...790....7J}
{Jiang}, C.~Y., {Jing}, Y.~P., \& {Han}, J. 2014, \apj, 790, 7,
  \dodoi{10.1088/0004-637X/790/1/7}

\bibitem[{{Jiang} {et~al.}(2015){Jiang}, {Cole}, {Sawala}, \&
  {Frenk}}]{2015MNRAS.448.1674J}
{Jiang}, L., {Cole}, S., {Sawala}, T., \& {Frenk}, C.~S. 2015, \mnras, 448,
  1674, \dodoi{10.1093/mnras/stv053}

\bibitem[{{Jing}(2019)}]{2019SCPMA..6219511J}
{Jing}, Y. 2019, Science China Physics, Mechanics, and Astronomy, 62, 19511,
  \dodoi{10.1007/s11433-018-9286-x}

\bibitem[{{Jing} \& {Suto}(2000)}]{2000ApJ...529L..69J}
{Jing}, Y.~P., \& {Suto}, Y. 2000, \apjl, 529, L69, \dodoi{10.1086/312463}

\bibitem[{{Jing} \& {Suto}(2002)}]{2002ApJ...574..538J}
---. 2002, \apj, 574, 538, \dodoi{10.1086/341065}

\bibitem[{{Kauffmann} {et~al.}(1993){Kauffmann}, {White}, \&
  {Guiderdoni}}]{1993MNRAS.264..201K}
{Kauffmann}, G., {White}, S.~D.~M., \& {Guiderdoni}, B. 1993, \mnras, 264, 201,
  \dodoi{10.1093/mnras/264.1.201}

\bibitem[{{Kazantzidis} {et~al.}(2004){Kazantzidis}, {Mayer}, {Mastropietro},
  {Diemand}, {Stadel}, \& {Moore}}]{2004ApJ...608..663K}
{Kazantzidis}, S., {Mayer}, L., {Mastropietro}, C., {et~al.} 2004, \apj, 608,
  663, \dodoi{10.1086/420840}

\bibitem[{{Kitzbichler} \& {White}(2008)}]{2008MNRAS.391.1489K}
{Kitzbichler}, M.~G., \& {White}, S.~D.~M. 2008, \mnras, 391, 1489,
  \dodoi{10.1111/j.1365-2966.2008.13873.x}

\bibitem[{{Lacey} \& {Cole}(1993)}]{1993MNRAS.262..627L}
{Lacey}, C., \& {Cole}, S. 1993, \mnras, 262, 627,
  \dodoi{10.1093/mnras/262.3.627}

\bibitem[{{Malhan} {et~al.}(2018){Malhan}, {Ibata}, \&
  {Martin}}]{2018MNRAS.481.3442M}
{Malhan}, K., {Ibata}, R.~A., \& {Martin}, N.~F. 2018, \mnras, 481, 3442,
  \dodoi{10.1093/mnras/sty2474}

\bibitem[{{Marinacci} {et~al.}(2018){Marinacci}, {Vogelsberger}, {Pakmor},
  {Torrey}, {Springel}, {Hernquist}, {Nelson}, {Weinberger}, {Pillepich},
  {Naiman}, \& {Genel}}]{2018MNRAS.480.5113M}
{Marinacci}, F., {Vogelsberger}, M., {Pakmor}, R., {et~al.} 2018, \mnras, 480,
  5113, \dodoi{10.1093/mnras/sty2206}

\bibitem[{{Mo} {et~al.}(2010){Mo}, {van den Bosch}, \&
  {White}}]{2010gfe..book.....M}
{Mo}, H., {van den Bosch}, F.~C., \& {White}, S. 2010, {Galaxy Formation and
  Evolution} (Cambridge University Press)

\bibitem[{{Naiman} {et~al.}(2018){Naiman}, {Pillepich}, {Springel},
  {Ramirez-Ruiz}, {Torrey}, {Vogelsberger}, {Pakmor}, {Nelson}, {Marinacci},
  {Hernquist}, {Weinberger}, \& {Genel}}]{2018MNRAS.477.1206N}
{Naiman}, J.~P., {Pillepich}, A., {Springel}, V., {et~al.} 2018, \mnras, 477,
  1206, \dodoi{10.1093/mnras/sty618}

\bibitem[{{Navarro} {et~al.}(1995){Navarro}, {Frenk}, \&
  {White}}]{1995MNRAS.275...56N}
{Navarro}, J.~F., {Frenk}, C.~S., \& {White}, S. D.~M. 1995, \mnras, 275, 56,
  \dodoi{10.1093/mnras/275.1.56}

\bibitem[{{Navarro} {et~al.}(1997){Navarro}, {Frenk}, \&
  {White}}]{1997ApJ...490..493N}
---. 1997, \apj, 490, 493, \dodoi{10.1086/304888}

\bibitem[{{Nelson} {et~al.}(2018){Nelson}, {Pillepich}, {Springel},
  {Weinberger}, {Hernquist}, {Pakmor}, {Genel}, {Torrey}, {Vogelsberger},
  {Kauffmann}, {Marinacci}, \& {Naiman}}]{2018MNRAS.475..624N}
{Nelson}, D., {Pillepich}, A., {Springel}, V., {et~al.} 2018, \mnras, 475, 624,
  \dodoi{10.1093/mnras/stx3040}

\bibitem[{{Nelson} {et~al.}(2019){Nelson}, {Springel}, {Pillepich},
  {Rodriguez-Gomez}, {Torrey}, {Genel}, {Vogelsberger}, {Pakmor}, {Marinacci},
  {Weinberger}, {Kelley}, {Lovell}, {Diemer}, \&
  {Hernquist}}]{2019ComAC...6....2N}
{Nelson}, D., {Springel}, V., {Pillepich}, A., {et~al.} 2019, Computational
  Astrophysics and Cosmology, 6, 2, \dodoi{10.1186/s40668-019-0028-x}

\bibitem[{{Pillepich} {et~al.}(2018{\natexlab{a}}){Pillepich}, {Nelson},
  {Hernquist}, {Springel}, {Pakmor}, {Torrey}, {Weinberger}, {Genel}, {Naiman},
  {Marinacci}, \& {Vogelsberger}}]{2018MNRAS.475..648P}
{Pillepich}, A., {Nelson}, D., {Hernquist}, L., {et~al.} 2018{\natexlab{a}},
  \mnras, 475, 648, \dodoi{10.1093/mnras/stx3112}

\bibitem[{{Pillepich} {et~al.}(2018{\natexlab{b}}){Pillepich}, {Springel},
  {Nelson}, {Genel}, {Naiman}, {Pakmor}, {Hernquist}, {Torrey}, {Vogelsberger},
  {Weinberger}, \& {Marinacci}}]{2018MNRAS.473.4077P}
{Pillepich}, A., {Springel}, V., {Nelson}, D., {et~al.} 2018{\natexlab{b}},
  \mnras, 473, 4077, \dodoi{10.1093/mnras/stx2656}

\bibitem[{{Planck Collaboration} {et~al.}(2016){Planck Collaboration}, {Ade},
  {Aghanim}, {Arnaud}, {Ashdown}, {Aumont}, {Baccigalupi}, {Banday},
  {Barreiro}, {Bartlett}, {Bartolo}, {Battaner}, {Battye}, {Benabed},
  {Beno{\^\i}t}, {Benoit-L{\'e}vy}, {Bernard}, {Bersanelli}, {Bielewicz},
  {Bock}, {Bonaldi}, {Bonavera}, {Bond}, {Borrill}, {Bouchet}, {Boulanger},
  {Bucher}, {Burigana}, {Butler}, {Calabrese}, {Cardoso}, {Catalano},
  {Challinor}, {Chamballu}, {Chary}, {Chiang}, {Chluba}, {Christensen},
  {Church}, {Clements}, {Colombi}, {Colombo}, {Combet}, {Coulais}, {Crill},
  {Curto}, {Cuttaia}, {Danese}, {Davies}, {Davis}, {de Bernardis}, {de Rosa},
  {de Zotti}, {Delabrouille}, {D{\'e}sert}, {Di Valentino}, {Dickinson},
  {Diego}, {Dolag}, {Dole}, {Donzelli}, {Dor{\'e}}, {Douspis}, {Ducout},
  {Dunkley}, {Dupac}, {Efstathiou}, {Elsner}, {En{\ss}lin}, {Eriksen},
  {Farhang}, {Fergusson}, {Finelli}, {Forni}, {Frailis}, {Fraisse},
  {Franceschi}, {Frejsel}, {Galeotta}, {Galli}, {Ganga}, {Gauthier}, {Gerbino},
  {Ghosh}, {Giard}, {Giraud-H{\'e}raud}, {Giusarma}, {Gjerl{\o}w},
  {Gonz{\'a}lez-Nuevo}, {G{\'o}rski}, {Gratton}, {Gregorio}, {Gruppuso},
  {Gudmundsson}, {Hamann}, {Hansen}, {Hanson}, {Harrison}, {Helou},
  {Henrot-Versill{\'e}}, {Hern{\'a}ndez-Monteagudo}, {Herranz}, {Hildebrandt},
  {Hivon}, {Hobson}, {Holmes}, {Hornstrup}, {Hovest}, {Huang}, {Huffenberger},
  {Hurier}, {Jaffe}, {Jaffe}, {Jones}, {Juvela}, {Keih{\"a}nen}, {Keskitalo},
  {Kisner}, {Kneissl}, {Knoche}, {Knox}, {Kunz}, {Kurki-Suonio}, {Lagache},
  {L{\"a}hteenm{\"a}ki}, {Lamarre}, {Lasenby}, {Lattanzi}, {Lawrence}, {Leahy},
  {Leonardi}, {Lesgourgues}, {Levrier}, {Lewis}, {Liguori}, {Lilje},
  {Linden-V{\o}rnle}, {L{\'o}pez-Caniego}, {Lubin}, {Mac{\'\i}as-P{\'e}rez},
  {Maggio}, {Maino}, {Mandolesi}, {Mangilli}, {Marchini}, {Maris}, {Martin},
  {Martinelli}, {Mart{\'\i}nez-Gonz{\'a}lez}, {Masi}, {Matarrese}, {McGehee},
  {Meinhold}, {Melchiorri}, {Melin}, {Mendes}, {Mennella}, {Migliaccio},
  {Millea}, {Mitra}, {Miville-Desch{\^e}nes}, {Moneti}, {Montier}, {Morgante},
  {Mortlock}, {Moss}, {Munshi}, {Murphy}, {Naselsky}, {Nati}, {Natoli},
  {Netterfield}, {N{\o}rgaard-Nielsen}, {Noviello}, {Novikov}, {Novikov},
  {Oxborrow}, {Paci}, {Pagano}, {Pajot}, {Paladini}, {Paoletti}, {Partridge},
  {Pasian}, {Patanchon}, {Pearson}, {Perdereau}, {Perotto}, {Perrotta},
  {Pettorino}, {Piacentini}, {Piat}, {Pierpaoli}, {Pietrobon}, {Plaszczynski},
  {Pointecouteau}, {Polenta}, {Popa}, {Pratt}, {Pr{\'e}zeau}, {Prunet},
  {Puget}, {Rachen}, {Reach}, {Rebolo}, {Reinecke}, {Remazeilles}, {Renault},
  {Renzi}, {Ristorcelli}, {Rocha}, {Rosset}, {Rossetti}, {Roudier},
  {Rouill{\'e} d'Orfeuil}, {Rowan-Robinson}, {Rubi{\~n}o-Mart{\'\i}n},
  {Rusholme}, {Said}, {Salvatelli}, {Salvati}, {Sandri}, {Santos},
  {Savelainen}, {Savini}, {Scott}, {Seiffert}, {Serra}, {Shellard}, {Spencer},
  {Spinelli}, {Stolyarov}, {Stompor}, {Sudiwala}, {Sunyaev}, {Sutton},
  {Suur-Uski}, {Sygnet}, {Tauber}, {Terenzi}, {Toffolatti}, {Tomasi},
  {Tristram}, {Trombetti}, {Tucci}, {Tuovinen}, {T{\"u}rler}, {Umana},
  {Valenziano}, {Valiviita}, {Van Tent}, {Vielva}, {Villa}, {Wade}, {Wandelt},
  {Wehus}, {White}, {White}, {Wilkinson}, {Yvon}, {Zacchei}, \&
  {Zonca}}]{2016A&A...594A..13P}
{Planck Collaboration}, {Ade}, P.~A.~R., {Aghanim}, N., {et~al.} 2016, \aap,
  594, A13, \dodoi{10.1051/0004-6361/201525830}

\bibitem[{{Poulton} {et~al.}(2021){Poulton}, {Power}, {Robotham}, {Elahi}, \&
  {Lagos}}]{2021MNRAS.501.2810P}
{Poulton}, R.~J.~J., {Power}, C., {Robotham}, A.~S.~G., {Elahi}, P.~J., \&
  {Lagos}, C.~D.~P. 2021, \mnras, 501, 2810, \dodoi{10.1093/mnras/staa3247}

\bibitem[{{Press} \& {Schechter}(1974)}]{1974ApJ...187..425P}
{Press}, W.~H., \& {Schechter}, P. 1974, \apj, 187, 425, \dodoi{10.1086/152650}

\bibitem[{{Rodriguez-Gomez} {et~al.}(2015){Rodriguez-Gomez}, {Genel},
  {Vogelsberger}, {Sijacki}, {Pillepich}, {Sales}, {Torrey}, {Snyder},
  {Nelson}, {Springel}, {Ma}, \& {Hernquist}}]{2015MNRAS.449...49R}
{Rodriguez-Gomez}, V., {Genel}, S., {Vogelsberger}, M., {et~al.} 2015, \mnras,
  449, 49, \dodoi{10.1093/mnras/stv264}

\bibitem[{{Rodriguez-Gomez} {et~al.}(2016){Rodriguez-Gomez}, {Pillepich},
  {Sales}, {Genel}, {Vogelsberger}, {Zhu}, {Wellons}, {Nelson}, {Torrey},
  {Springel}, {Ma}, \& {Hernquist}}]{2016MNRAS.458.2371R}
{Rodriguez-Gomez}, V., {Pillepich}, A., {Sales}, L.~V., {et~al.} 2016, \mnras,
  458, 2371, \dodoi{10.1093/mnras/stw456}

\bibitem[{{Samuel} {et~al.}(2020){Samuel}, {Wetzel}, {Tollerud},
  {Garrison-Kimmel}, {Loebman}, {El-Badry}, {Hopkins}, {Boylan-Kolchin},
  {Faucher-Gigu{\`e}re}, {Bullock}, {Benincasa}, \&
  {Bailin}}]{2020MNRAS.491.1471S}
{Samuel}, J., {Wetzel}, A., {Tollerud}, E., {et~al.} 2020, \mnras, 491, 1471,
  \dodoi{10.1093/mnras/stz3054}

\bibitem[{{Simha} \& {Cole}(2017)}]{2017MNRAS.472.1392S}
{Simha}, V., \& {Cole}, S. 2017, \mnras, 472, 1392,
  \dodoi{10.1093/mnras/stx1942}

\bibitem[{{Springel}(2010)}]{2010MNRAS.401..791S}
{Springel}, V. 2010, \mnras, 401, 791, \dodoi{10.1111/j.1365-2966.2009.15715.x}

\bibitem[{{Springel} {et~al.}(2001){Springel}, {White}, {Tormen}, \&
  {Kauffmann}}]{2001MNRAS.328..726S}
{Springel}, V., {White}, S. D.~M., {Tormen}, G., \& {Kauffmann}, G. 2001,
  \mnras, 328, 726, \dodoi{10.1046/j.1365-8711.2001.04912.x}

\bibitem[{{Springel} {et~al.}(2018){Springel}, {Pakmor}, {Pillepich},
  {Weinberger}, {Nelson}, {Hernquist}, {Vogelsberger}, {Genel}, {Torrey},
  {Marinacci}, \& {Naiman}}]{2018MNRAS.475..676S}
{Springel}, V., {Pakmor}, R., {Pillepich}, A., {et~al.} 2018, \mnras, 475, 676,
  \dodoi{10.1093/mnras/stx3304}

\bibitem[{{St{\"u}cker} {et~al.}(2023){St{\"u}cker}, {Ogiya}, {Angulo},
  {Aguirre-Santaella}, \& {S{\'a}nchez-Conde}}]{2023MNRAS.521.4432S}
{St{\"u}cker}, J., {Ogiya}, G., {Angulo}, R.~E., {Aguirre-Santaella}, A., \&
  {S{\'a}nchez-Conde}, M.~A. 2023, \mnras, 521, 4432,
  \dodoi{10.1093/mnras/stad844}

\bibitem[{{Taffoni} {et~al.}(2003){Taffoni}, {Mayer}, {Colpi}, \&
  {Governato}}]{2003MNRAS.341..434T}
{Taffoni}, G., {Mayer}, L., {Colpi}, M., \& {Governato}, F. 2003, \mnras, 341,
  434, \dodoi{10.1046/j.1365-8711.2003.06395.x}

\bibitem[{{Taylor} \& {Babul}(2001)}]{2001ApJ...559..716T}
{Taylor}, J.~E., \& {Babul}, A. 2001, \apj, 559, 716, \dodoi{10.1086/322276}

\bibitem[{{Toomre} \& {Toomre}(1972)}]{1972ApJ...178..623T}
{Toomre}, A., \& {Toomre}, J. 1972, \apj, 178, 623, \dodoi{10.1086/151823}

\bibitem[{{Toth} \& {Ostriker}(1992)}]{1992ApJ...389....5T}
{Toth}, G., \& {Ostriker}, J.~P. 1992, \apj, 389, 5, \dodoi{10.1086/171185}

\bibitem[{{van den Bosch} \& {Ogiya}(2018)}]{2018MNRAS.475.4066V}
{van den Bosch}, F.~C., \& {Ogiya}, G. 2018, \mnras, 475, 4066,
  \dodoi{10.1093/mnras/sty084}

\bibitem[{{van den Bosch} {et~al.}(2018){van den Bosch}, {Ogiya}, {Hahn}, \&
  {Burkert}}]{2018MNRAS.474.3043V}
{van den Bosch}, F.~C., {Ogiya}, G., {Hahn}, O., \& {Burkert}, A. 2018, \mnras,
  474, 3043, \dodoi{10.1093/mnras/stx2956}

\bibitem[{{Velazquez} \& {White}(1999)}]{1999MNRAS.304..254V}
{Velazquez}, H., \& {White}, S. D.~M. 1999, \mnras, 304, 254,
  \dodoi{10.1046/j.1365-8711.1999.02354.x}

\bibitem[{{Wechsler} {et~al.}(2002){Wechsler}, {Bullock}, {Primack},
  {Kravtsov}, \& {Dekel}}]{2002ApJ...568...52W}
{Wechsler}, R.~H., {Bullock}, J.~S., {Primack}, J.~R., {Kravtsov}, A.~V., \&
  {Dekel}, A. 2002, \apj, 568, 52, \dodoi{10.1086/338765}

\bibitem[{{Weinberger} {et~al.}(2017){Weinberger}, {Springel}, {Hernquist},
  {Pillepich}, {Marinacci}, {Pakmor}, {Nelson}, {Genel}, {Vogelsberger},
  {Naiman}, \& {Torrey}}]{2017MNRAS.465.3291W}
{Weinberger}, R., {Springel}, V., {Hernquist}, L., {et~al.} 2017, \mnras, 465,
  3291, \dodoi{10.1093/mnras/stw2944}

\bibitem[{{Wetzel} \& {White}(2010)}]{2010MNRAS.403.1072W}
{Wetzel}, A.~R., \& {White}, M. 2010, \mnras, 403, 1072,
  \dodoi{10.1111/j.1365-2966.2009.16191.x}

\bibitem[{{White}(1976)}]{1976MNRAS.174...19W}
{White}, S.~D.~M. 1976, \mnras, 174, 19, \dodoi{10.1093/mnras/174.1.19}

\bibitem[{{White} \& {Frenk}(1991)}]{1991ApJ...379...52W}
{White}, S. D.~M., \& {Frenk}, C.~S. 1991, \apj, 379, 52,
  \dodoi{10.1086/170483}

\bibitem[{{White} \& {Rees}(1978)}]{1978MNRAS.183..341W}
{White}, S.~D.~M., \& {Rees}, M.~J. 1978, \mnras, 183, 341,
  \dodoi{10.1093/mnras/183.3.341}

\bibitem[{{Whitney} {et~al.}(2021){Whitney}, {Ferreira}, {Conselice}, \&
  {Duncan}}]{2021ApJ...919..139W}
{Whitney}, A., {Ferreira}, L., {Conselice}, C.~J., \& {Duncan}, K. 2021, \apj,
  919, 139, \dodoi{10.3847/1538-4357/ac1422}

\bibitem[{{Xu} {et~al.}(2024){Xu}, {Jing}, {Gao}, {Luo}, \&
  {Li}}]{2024ApJ...973..102X}
{Xu}, K., {Jing}, Y.~P., {Gao}, H., {Luo}, X., \& {Li}, M. 2024, \apj, 973,
  102, \dodoi{10.3847/1538-4357/ad6156}

\bibitem[{{Zhao} {et~al.}(2009){Zhao}, {Jing}, {Mo}, \&
  {B{\"o}rner}}]{2009ApJ...707..354Z}
{Zhao}, D.~H., {Jing}, Y.~P., {Mo}, H.~J., \& {B{\"o}rner}, G. 2009, \apj, 707,
  354, \dodoi{10.1088/0004-637X/707/1/354}

\bibitem[{{Zhao} {et~al.}(2003){Zhao}, {Mo}, {Jing}, \&
  {B{\"o}rner}}]{2003MNRAS.339...12Z}
{Zhao}, D.~H., {Mo}, H.~J., {Jing}, Y.~P., \& {B{\"o}rner}, G. 2003, \mnras,
  339, 12, \dodoi{10.1046/j.1365-8711.2003.06135.x}

\end{thebibliography}
\bibliographystyle{aasjournal}



\end{document}